\documentclass[preprint2]{aastex}
\usepackage{natbib, hyperref}
\usepackage{amssymb,amsmath}
\usepackage{longtable, float, subfigure}
\usepackage[T1]{fontenc}
\usepackage[latin1]{inputenc}

\def\kms{km s$^{-1}$}
\def\et{{et~al.}}
\def\ha{H$\alpha$}
\def\arcmin{\ifmmode^\prime\;\else$^\prime$\fi}
\def\arcsec{\ifmmode^{\prime\prime}\;\else$^{\prime\prime}$\fi}
\def\deg{\ifmmode^\circ\;\else$^\circ$\fi}

\shorttitle{Stellar and Gas Kinematics of NGC 1569}
\shortauthors{Johnson et al.\ }

\begin{document}


\title{The Stellar and Gas Kinematics of the LITTLE THINGS Dwarf Irregular Galaxy NGC 1569}


\author {Megan Johnson\altaffilmark{1,2,3,10},
Deidre A. Hunter\altaffilmark{2,10},
Se-Heon Oh\altaffilmark{4,5},  
Hong-Xin Zhang\altaffilmark{2,6},
Bruce Elmegreen\altaffilmark{7},
Elias Brinks\altaffilmark{8},
Erik Tollerud\altaffilmark{9},
Kimberly Herrmann\altaffilmark{2}
}



\altaffiltext{1}{National Radio Astronomy Observatory, PO Box 2, Green Bank, WV 24944; mjohnson@nrao.edu}
\altaffiltext{2}{Lowell Observatory, 1400 West Mars Hill Road, Flagstaff, AZ 86001; dah@lowell.edu, hxzhang@lowell.edu, herrmann@lowell.edu}
\altaffiltext{3}{Georgia State University, 1 Park Place South SE, Suite 700, Atlanta, GA 30303-2911} 
\altaffiltext{4}{International Centre for Radio Astronomy Research (ICRAR), University of Western Australia, 35 Stirling Highway, Crawley, WA 6009, Australia; se-heon.oh@uwa.edu.au}
\altaffiltext{5}{ARC Centre of Excellence for All-sky Astrophysics (CAASTRO)} 
\altaffiltext{6}{Purple Mountain Observatory, Chinese Academy of Sciences}
\altaffiltext{7}{IBM T.J. Watson Research Center,
1101 Kitchawan Road,
Yorktown Hts., NY 10598; bge@us.ibm.com}
\altaffiltext{8}{Centre for Astrophysics Research, University of Hertfordshire, College Lane, Hatfield, AL10 9AB, UK; E.Brinks@herts.ac.uk}
\altaffiltext{9}{Center For Cosmology,
Department of Physics and Astronomy,
4129 Frederick Reines Hall,
University of California, Irvine;
etolleru@uci.edu}
\altaffiltext{10}{Visiting Astronomer, Kitt Peak National Observatory}

\normalsize
\begin{abstract}
In order to understand the formation and evolution of Magellanic-type dwarf irregular (dIm) galaxies, one needs to understand their three-dimensional structure.
We present measurements of the stellar velocity dispersion in NGC 1569, a nearby post-starburst dIm galaxy.  The stellar vertical velocity dispersion, $\sigma_{\rm z}$, coupled with the maximum rotational velocity derived from \ion{H}{1} observations, $V_{\rm max}$, gives a measure of how kinematically hot the galaxy is, and, therefore, indicates its structure. We conclude that the stars in NGC 1569 are in a thick disk with a $V_{\rm max}$/$\sigma_{\rm z}$ = 2.4 $\pm$ 0.7.  In addition to the structure, we analyze the ionized gas kinematics from \ion{O}{3} observations along the morphological major axis.  These data show evidence for outflow from the inner starburst region and a potential expanding shell near supermassive star cluster (SSC) A.  When compared to the stellar kinematics, the velocity dispersion of the stars increase in the region of SSC A supporting the hypothesis of an expanding shell.   
The stellar kinematics closely follow the motion of the gas.  Analysis of high resolution \ion{H}{1} data clearly reveals the presence of an \ion{H}{1} cloud that appears to be impacting the eastern edge of NGC 1569.  Also, an ultra-dense \ion{H}{1} cloud can be seen extending to the west of the impacting \ion{H}{1} cloud.  This dense cloud is likely the remains of a dense \ion{H}{1} bridge that extended through what is now the central starburst area. The impacting \ion{H}{1} cloud was the catalyst for the starburst, thus turning the dense gas into stars over a short timescale, $\sim$ 1 Gyr.  We performed a careful study of the spectral energy distribution using infrared, optical, and ultraviolet photometry producing a state-of-the-art mass model for the stellar disk. This mass modeling shows that stars dominate the gravitational potential in the inner 1 kpc.  The dynamical mass of NGC 1569, derived from $V_{\rm max}$, shows that the disk may be dark matter deficient in the inner region, although, when compared to the expected virial mass determined from halo abundance matching techniques, the dark matter profile seems to agree with the observed mass profile at a radius of 2.2 kpc. 
\end{abstract}


\keywords{galaxies: individual (NGC 1569) --- galaxies: dwarf galaxies --- galaxies: starburst --- galaxies: kinematics --- galaxies: spectroscopy}



\section{Introduction} \label{sec:1}

Dwarf galaxies are the most common type of galaxy and
are proposed as the building blocks of larger galaxies
in the early universe.
Yet, even the basic structure of Magellanic-type dwarf irregular (dIm) galaxies remains controversial.
Studies of the distributions of projected minor-to-major axis ratios
$b/a$ have been interpreted to mean that dIm galaxies are thick
disks with intrinsic ratios $(b/a)_0$ 1.5--3 times that of spirals
\citep{hod66, van88, sta92}.
Yet others have interpreted the distribution as evidence that dIm galaxies
are triaxial 
\citep{bin95},
only a little less spherical than dwarf ellipticals (dE) \citep{sun98}
Clearly, statistical analyses of axis ratios
have not been adequate to reveal the true shapes of dIm galaxies.

However, the {\it stellar} kinematics of a single galaxy, defines
its intrinsic shape. The ratio $V_{\rm max}/\sigma_z$ is a measure of how 
kinematically hot a system is and, therefore, reveals its structure.
Here, $V_{\rm max}$ is the maximum speed of rotation of the galaxy and $\sigma_z$ is the intrinsic stellar velocity dispersion perpendicular to the disk. 
Elliptical and dEs have $V_{\rm max}/\sigma_z<1$ 
\citep[Figure 4-6 in][]{bin87, ped02}
whereas spirals have $V_{\rm max}/\sigma_z>1$, usually 2--5
\citep{bot93, veg01}

We study the stellar and gas kinematics of NGC 1569, with the goal of determining the three-dimensional shape by measuring $V_{\rm max}/\sigma_z$.  We also aim to understand its recent starburst from an analysis of the kinematics.
The stellar kinematics are derived from long-slit spectroscopy of integrated light, in the sense that the stars are unresolved, and the high-resolution \ion{H}{1} kinematics are from The \ion{H}{1} Nearby Galaxy Survey \citep[THINGS;]{wal08}.   Additionally, we discuss a new method of imaging the high-resolution \ion{H}{1} data cube and apply a double Gaussian decomposition of the \ion{H}{1} line profiles, a method developed by \citet{oh08}, in order to separate ordered motions from non-ordered motions. 


NGC 1569 is a nearby post-starburst dIm galaxy.  It has three supermassive star clusters (SSCs) that formed at the end of the most recent episode of intense star formation that concluded approximately 10 million years ago \citep[][and references therein]{hun00}.
This dIm galaxy has complex gas kinematics. \citet{sti98} used the Westerbork Synthesis Radio Telescope (WSRT) to obtain high-sensitivity \ion{H}{1} emission maps of NGC 1569 and show the presence of a strong non-circular motion (NCM) \ion{H}{1} cloud \citep{sti02} and a possible \ion{H}{1} companion and bridge that are to the east of the disk \citep{sti98, muh05}.
\citet{muh05} also observed these features and suggest that they are being accreted onto NGC 1569.  In the center of the galaxy, energy from supernova explosions from the most massive stars in SSC A has blown a hole in the \ion{H}{1} gas \citep{isr90}.  There is X-ray emission to the north and south of the galaxy, perpendicular to the disk, from the outflow of hot, ionized gas driven by the starburst \citep{ott05a}, and strong magnetic fields appear to be channeling the gas outflow \citep{kep10}.  There is an \ha\ chimney extending to the north showing the escape route of the gas \citep{hun93, muh05}, which has been suggested as evidence for blowout \citep{wes08}.  

Figure \ref{fig:cartoon} outlines the major \ion{H}{1} features, including the suggested \ion{H}{1} companion and bridge \citep{sti02}, the NCM \ion{H}{1} cloud \citep{sti02}, an ultra-dense (UD) \ion{H}{1} cloud described below, and extended \ion{H}{1} emission imaged in this work.  This figure also identifies the \ha\ chimney and `western arm' \citep{hod74}, shows the location of SSCs A and B, and marks the location, size and position of the optical slit used to determine the stellar kinematics along the major axis of NGC 1569 discussed in this work.  The sense of rotation of the stellar and gas disks are marked by the white curved arrows and the near and far sides of the disks compared to Earth are also labeled.  The relative velocities with respect to the systemic velocity, $V_{\rm sys}$, of the \ion{H}{1} clouds and features are also shown in Figure \ref{fig:cartoon}.

Table \ref{tab:1} gives some global parameters of NGC 1569.  Note that the distance \citep[3.36 $\pm$ 0.20 Mpc;][]{gro08} places NGC 1569 comfortably within the IC 342 galaxy group, which has an average distance of 3.35 $\pm$ 0.09 Mpc \citep{kar05}.  Having a group association raises the possibility of NGC 1569 interacting or merging with other galaxies in the IC 342 galaxy group, potentially explaining its unusual characteristics.  
NGC 1569 is, in fact, a complex system with complicated kinematics, and we will present evidence here and in a forthcoming paper (Johnson, in prep) of an interaction or a past merger that has created the dIm galaxy we observe today.  


This work is part of a large collaborative effort called Local Irregulars That Trace Luminosity Extremes, The \ion{H}{1} Nearby Galaxy Survey \citep[LITTLE THINGS;][]{hun12}.  One of the goals of the LITTLE THINGS survey is to understand how dIm galaxies form stars.  LITTLE THINGS is a multi-wavelength study of 37 dIm and 4 Blue Compact Dwarf (BCD) galaxies.  LITTLE THINGS was granted close to 376 hours in 2007 and 2008 on the NRAO\footnote{National Radio Astronomy Observatory (NRAO) is a facility of the National Science Foundation operated under cooperative
agreement by Associated Universities, Inc.} Very Large Array (VLA) in the B-, C-, and D-array configurations in order to obtain deep \ion{H}{1} maps for 21 galaxies.  The other 20 galaxies are from the VLA archives.  In addition to the \ion{H}{1} data, LITTLE THINGS has extensive H$\alpha$ and optical photometry for each dIm galaxy from previous studies \citep{hun04, hun06} and has also collected {\it GALEX} ultraviolet images for most and near- and mid-infrared data (ground based $JHK$ and {\it Spitzer}) for a subsample.

Our paper is structured as follows:  Section 2 describes the observations and data reductions for the optical spectroscopy and the \ion{H}{1} data used in this study.  Section 3 discusses our data analysis; Section 4 outlines our observational results; Section 5 describes our mass modeling; Section 6 is our discussion; Section 7 is a summary of our results.

\section{Data} \label{sec:2}


	\subsection{Stellar Optical Spectroscopy} \label{sec:2.1}

		\subsubsection{Observations} \label{sec:2.1.1}

We observed NGC 1569 with the Kitt Peak National Observatory (KPNO\footnote{The Kitt Peak National Observatory is operated by the National Optical Astronomy Observatory (NOAO), which is operated by the Association of Universities for Research in Astronomy (AURA) under cooperative agreement with the National Science Foundation.}) Mayall 4-meter telescope with the Echelle spectrograph for three nights in February 2008 and five nights in November 2008. The sky conditions for the February 2008 observing run were moderate with only a few cirrus clouds interfering sporadically throughout the run.  The moon was in a waning crescent phase and was separated by more than 180$\arcdeg$ from NGC 1569.  The weather conditions for the November 2008 observing run were superb with nearly photometric conditions throughout the five nights, however, the moon passed through third quarter during the second night of our run but was separated from NGC 1569 by more than 60$\arcdeg$.

Table \ref{tab:2} lists the observation log.  In order to place the slit, we identified the center of the galaxy and position angle (PA) of the major axis from iso-intensity contours on our V-band image.  We observed at four PAs, all centered on the morphological center of the galaxy:  the major axis (PA = 121$\arcdeg$), the minor axis (PA = 31$\arcdeg$) and $\pm$45$\arcdeg$ from the major axis.  Figure \ref{fig:46} shows the 
{\it HST}\footnote{{\it Hubble Space Telescope} ({\it HST)}, is operated by NASA and ESA at the Space Telescope Science Institute (STScI).  STScI is operated by the Association of Universities for Research in Astronomy, Inc., under NASA contract NAS 5-26555.} V-band, \ion{O}{3} $\lambda$5007\AA, H$\beta$ and H$\alpha$ filter composite image of NGC 1569 taken with the Advanced Camera for Surveys and the Wide Field Planetary Camera 2 instruments (bottom panel) with the slit positions superimposed.  

We converted the Echelle into a two-dimensional, long-slit spectrograph by replacing the cross-disperser with a mirror flat.  This permitted the isolation of a single order and produced a 3$\arcmin$ long slit.  
In order to maximize throughput, while maintaining high spectral resolution, we opened the slit to a width of 2$\farcs5$.  For the stellar kinematic data, we targeted the prominent Mg Ib stellar absorption features at wavelengths 5183.6 \AA, 5172.7 \AA, and 5167.3 \AA, by isolating order 10 for the 316-63$\arcdeg$ Echelle using the post-slit filter KP 1433.  This filter has a central wavelength of 5204 \AA\ and a full-width half-maximum (FWHM) of 276 \AA. The free spectral range of the Echelle was 517 \AA\ centered at 5170 \AA, but, for the Tektronics 2048 x 2048 CCD that we used, only 286 \AA\ fit across the chip.  We used the ``fast UV'' camera along with pre-slit filter GG 495.  
The prominent Mg Ib stellar absorption features were easily detected.  There are no telluric features in the spectral region we observed.

For our spectral range, this setup provided a FWHM $=$ 0.38 \AA\ as measured from the central lines of Th-Ar comparison lamp exposures, giving a velocity resolution of $\sim$22 \kms.  This corresponds to a velocity dispersion, $\sigma$, of $\sim$9 km s$^{-1}$.  We achieved an angular pixel scale of about 1$\farcs2$ pix$^{-1}$, after binning by 2 pixels in the cross-dispersion direction, and a final spectral pixel scale of 0.14 \AA\ pix$^{-1}$.  
  
For each observing run, we moved a single radial velocity standard star in 30$\arcsec$ steps along the slit and used these data to trace the cross-dispersion curvature of the slit on the CCD. 
We averaged 10 dome flats from each night to determine pixel-to-pixel variations in the detector.  
Also, we took twilight sky exposures that were used to correct for illumination variations.  The rotator on the Echelle was set to the desired PA before observations began, and we generally used one PA during a night.  

Each night we observed a series of radial velocity standard stars.  These Milky Way stars were selected to have a range of spectral types (F, G, and K) and luminosity classes (III - V)
to provide a varied spectral stellar library for the cross-correlation analysis.  Table \ref{tab:3} lists the standard stars used in the cross-correlation.  We observed some of the same stars from night to night and run to run to minimize and identify systematic offsets.  There were no trends identified over our narrow wavelength region between the radial velocity standard star spectral types or classes.

For NGC 1569, we took 1800 s exposures after first verifying the telescope pointing by offsetting from a nearby foreground star before each exposure.  To ensure proper sky subtraction after each galaxy exposure, we observed a nearby dark region of sky for 1800 s.  In some cases, the sky exposures were scaled by a constant to correct for differing sky conditions.  Following every star, galaxy, or sky exposure, we took Th-Ar comparison lamp exposures in order to acquire accurate wavelength calibration as a function of time and position. For each PA, we took five to seven exposures on the galaxy, which we co-added for a total integration time of 2.5 -- 3.5 hours, during a single night.  

		\subsubsection{Data Reductions} \label{sec:2.1.2}
		
All data reductions were accomplished using the Image Reduction and Analysis Facility (IRAF\footnote{IRAF is distributed by the National Optical Astronomy Observatory.}).  
We fit the overscan to all images in order to remove the electronic pedestal, and corrected for pixel-to-pixel variations using dome flats taken each night ({\sc ccdproc}).  
We trimmed the two-dimensional flats so that the steep drop-off of light towards the edges would not complicate the normalization function. 

The wavelength solution was modeled by fitting the comparison lamp exposures independently for each spectrum each night.  For a given night, the wavelength solutions were stable to within a tenth of an angstrom.  Because we observed comparison lamp exposures before every spectrum, our wavelength solutions were robust.  Thus, we were able to place all images on the same wavelength scale for each observing run ({\sc longslit, identify, reidentify, fitcoords}). We used the radial velocity standard star that was moved spatially along the slit to map out the distortion along rows of constant position, and then, we used the comparison lamp exposures to trace the distortion along columns of constant wavelength.  These fitting procedures produced coordinate maps that were then applied to every respective object exposure ({\sc transform}).  
Next, we applied a slit function that corrected for the minor illumination drop off at the edges of the slit in the cross-dispersion direction.  This was achieved by combining the twilight flats from a single night and fitting a two-dimensional function to the illumination pattern along bins of summed columns in the cross-dispersion direction ({\sc illum}).  

Before combining the spectra, we ensured proper galaxy alignment by over-plotting the spectra and matching the peak flux in individual images.  If necessary, a pixel shift was applied to align peaks.  Galaxy images from a single night with the same PA were combined into a single spectrum with an algorithm to reject cosmic rays ({\sc imcombine}).  The same combining process, without the shifting procedure, was applied to the sky images.  
Finally, we subtracted the co-added sky image from the co-added galaxy image while taking care to scale the combined sky image, when needed, to correct for differing exposure times and sky conditions. Figure \ref{fig:2} shows the co-added galaxy spectrum of the major axis of NGC 1569 before and after sky subtraction.

	\subsection{Ionized Gas Optical Spectroscopy} \label{sec:2.2}

		\subsubsection{Observations and Data Reductions}  \label{sec:2.2.1}
		
A different post-slit filter (KP 1547) was used for the ionized gas, but otherwise, the telescope setup was unchanged.  The KP 1547 filter has a central wavelength of about 5015 \AA\ and a FWHM of 27 \AA, and was used for targeting the \ion{O}{3} $\lambda5007$ \AA\ emission feature.  We observed both the major and minor axes.  We took dome flats, twilight flats, and stepped one radial velocity standard star along the slit spatially to correct for the slit distortion and to find the center of the slit in the same fashion as for the other filter, described previously. 

Additionally, we took Th-Ar comparison lamp exposures after all object images for accurate wavelength calibration.  We observed three galaxy and three sky positions for the major axis PA, and three galaxy positions for the minor axis.  However, due to threatening clouds during the course of the night, we were only able to observe one sky image for the minor axis.  The sky image was scaled to compensate for the large difference in exposure time.  Despite the difficulties, the brightness of the \ion{O}{3} emission line made it possible to obtain high signal-to-noise emission line spectra.  We followed the same data reduction procedure as for the stellar spectral data.  Figure \ref{fig:3} shows the fully reduced, two-dimensional, co-added, \ion{O}{3} $\lambda5007$ \AA\ emission spectrum for the major and minor axes of NGC 1569.  
		
	\subsection{\ion{H}{1} Data}	\label{sec:2.3}
	
		
The LITTLE THINGS project retrieved the calibrated \ion{H}{1} line {\em uv--}data from THINGS but improved upon the quality of the THINGS data products by applying a multi--scale cleaning algorithm implemented in the AIPS task {\sc imagr} \citep[see, e.g.,][]{cor08, ric08, gre09}. 
Instead of approximating all structures as a sum of point sources, this algorithm iteratively fits Gaussians to four angular scales of 135$\arcsec$, 45$\arcsec$, 15$\arcsec$, and 0$\arcsec$, respectively.  The multi-scale cleaning algorithm produces a single data cube in which all observed angular frequencies are properly represented and with near uniform (Gaussian) noise statistics, which is a great advantage over standard cleaning methods.  A beam size of 7$\farcs71$ x 7$\farcs04$ (with a position angle of 16$\fdg5$) and a spectral resolution of 2.6 \kms\ were achieved.

In order to separate noise from real galaxy emission, we used the Natural-weighted cube and smoothed it to a 25\arcsec$\times$ 25\arcsec\ ({\sc convl}) beam.  We blanked this smoothed cube twice, first for all pixels
below $2.5~\sigma$ of the rms of the smoothed cube ({\sc blank}), and then again by hand for any additional emission that was clearly noise.  This final, blanked cube became the master for the cubes with smaller beam sizes and was used as a conditional transfer.
For a comprehensive review of our mapping procedure, please see Hunter et al.\ (submitted).  

The half-power beam width (HPBW) of the primary beam is 32$\arcmin$, but the largest angular scale of emission that is expected to be recovered by the interferometer is $\sim$15$\arcmin$, which is determined by the D-array.  
Any smooth structure larger than $\sim$15$\arcmin$ will be invisible to the interferometer. This was useful in removing foreground Milky Way HI emission, which has velocities that partly overlap with some of the velocities encountered in NGC 1569. In essence, the interferometer is blind for emission from the Milky Way, and will only register its small--scale structure, which is still considerably more extended than that of NGC 1569. The multi--scale cleaning algorithm was able to model this Milky Way emission sufficiently well so emission from NGC 1569 could be isolated from the rest using the blanking method described above.
 A more in depth discussion of Milky Way confusion with NGC 1569 is discussed in our upcoming paper describing the map made with the Green Bank Telescope (GBT) (Johnson, in preparation, hereafter Paper II).

Figure \ref{fig:47} shows the LITTLE THINGS integrated flux map. 
This image shows tenuous emission to the south and northeast of the galaxy that are imaged for the first time.  Although faint, this low--level, extended emission was confirmed in follow up observations with the Robert C. Byrd Green Bank Telescope (GBT) and will be discussed in Paper II.  

	\section{Analysis of Data} \label{sec:3}


	\subsection{Stellar Optical Spectroscopy} \label{sec:3.1}

		\subsubsection{Extracting One-dimensional Spectra} \label{sec:3.1.1}


We extracted one-dimensional stellar spectra from the two-dimensional spectra of NGC 1569 ({\sc scopy}).
It was necessary to sum over a series of consecutive rows in the cross-dispersion direction during spectral extraction in order to increase the signal-to-noise (S/N).  In the portion of the spectrum that contains the brightest light, from radii 0$\arcsec$ to -25$\arcsec$ (West of center) measured from the center of the galaxy, which is also where the SSCs reside (shown in Figure \ref{fig:29}), we summed over five rows, 6$\farcs1$ of the major axis.  For the regions of the galaxy that have a lower light concentration, from radii 17$\arcsec$ to 0$\arcsec$ (East of center), we summed over ten rows of the major axis, 12$\farcs2$.  
To avoid diluting the true stellar velocities and velocity dispersions, we never summed over more than 12$\farcs2$, which, at a distance of 3.36 Mpc \citep{gro08}, corresponds to 195 pc.   

Figure \ref{fig:4} shows three, one-dimensional galaxy spectra and one stellar template spectrum for comparison.  The top and third panels were the result of summing over 12$\farcs2$ in the cross-dispersion direction along the major axis of NGC 1569 and the second panel summed over 6$\farcs1$.  These three spectra show the range of S/N for our galaxy spectra and all of our spectra used in the CCM had a S/N above $\sim$ 5.  The bottom panel shows HD 132737, a K0 III template star for comparison.  All of the standard stars had S/N above $\sim$ 50, usually near 90, although this was difficult to measure due to the lack of line-free continuum (see bottom panel of Figure \ref{fig:4}).

We extracted one-dimensional spectra of the radial velocity standard stars using {\sc apall} (see, e.g., bottom panel of Figure \ref{fig:4}).  We identified the center of the peak emission in each star and fit the background using the spatial regions of the slit that contain no stellar data.  In some cases, we used the twilight sky exposures to extract a solar spectrum, which was used in the cross-correlation method (CCM) as a G2 V radial velocity standard star.  Table \ref{tab:2} identifies the nights that utilized these solar spectra by the $+$1 added to the number of standard star spectra in column 9.  Once one-dimensional spectra were extracted, we added into the header of each star the $V_{\rm helio}$ radial velocities from the United States Nautical Almanac (2008).  These spectra of radial velocity standard stars are the templates that were used in the CCM.
We fit a cubic spline function to the continuum in the spectra of the galaxy and the template stars ({\sc continuum}).  The slit width of 2$\farcs$5 may have caused small telescope tracking errors for some of the template stars because the seeing on most nights was  around 1$\farcs$2.  However, the exposure times for the standard stars were short, less than a minute for a large portion (see Table \ref{tab:3}), so tracking errors do not a play a large role in the total error budget. 

		\subsubsection{Cross-Correlation Method} \label{sec:3.1.2}

We cross-correlated the template stars with each other to test the CCM and to check the observed stars' velocities, which was within a few percent of the stars' known velocities in most cases ({\sc fxcor}).   
The one-dimensional spectra from the galaxy were cross-correlated with the stellar templates observed during the same night.  We only performed cross-correlations of the stars with stars and stars with galaxy spectra that were obtained on the same night in order to minimize the uncertainties in the observational setup, such as varying position angles. The heliocentric radial velocities and velocity dispersions of the stars in NGC 1569 were determined, for each spectrum, by fitting a single Gaussian curve to the peak of the CCM.  We did not explore the possibility of asymmetric Gaussian profiles as the quality of our data were not good enough for accurate analysis. We made the assumption that the stars in NGC 1569 are virialized and the Gaussian profiles are symmetric.  We excluded regions around emission line features in order to avoid internal galaxy emission or residuals due to atmospheric emission lines that were not completely removed during the sky subtraction step (see the top image of Figure \ref{fig:2}, which shows residual sky emission lines around 5196.8 \AA\ and 5199.1 \AA, and the top panel of Figure \ref{fig:4}).  

We cross-correlated each template star with each galaxy position and averaged the $V_{\rm helio}$ values by weighting them by the standard deviations of the different template spectra, thus determining the $V_{\rm helio}$ values at each galaxy radius.  By weighting the $V_{\rm helio}$ values in this manner, an ill-fitting template would not contribute as much to the resulting $V_{\rm helio}$ of the galaxy spectra as a good-fitting template.  Also, we observed no trend between stellar spectral types or classes when determining a ``good-fitting'' versus an ``ill-fitting'' template in the CCM; S/N dominated the quality of the fit.   

To determine the intrinsic velocity dispersions of the stars, $\sigma_{\rm obs}$, in NGC 1569 as a function of radius, we employ the definition of a Gaussian and correct for instrumental broadening using the following equation:  
\begin{equation}\label{eqn:sigma}
\sigma_{\rm obs} = \frac{\sqrt{{\rm FWHM_{obs}}^2 - {\rm FWHM_{instr}}^2}}{{2.35}} 
\end{equation}
We determine the inherent, average instrumental full-width half-maximum, FWHM$_{\rm instr}$  = 35 \kms, by cross-correlating the template stars from a single night with each other.  The intrinsic broadening of a single star should be unresolved, so the FWHM$_{\rm instr}$ of the cross-correlation of the template stars with each other represents the instrumental signature.  
Although the spectral types and classes of the radial velocity standard stars span a range, our usable wavelength region is only $\sim$ 150 \AA, thus cross-correlations of the stars between differing spectral types did not create variations in the resulting FWHM$_{\rm instr}$ or FWHM$_{\rm obs}$ above their respective standard deviations.


	\subsection{Ionized Gas Optical Spectroscopy}  \label{sec:3.2}
	
		\subsubsection{Extracting One-Dimensional Spectra}  \label{sec:3.2.1}

Due to the starburst nature of NGC 1569, many star forming regions lie throughout the observable disk and create bright \ion{O}{3} features seen along the cross-dispersion direction of these spectra.  Figure \ref{fig:3} shows the two-dimensional spectra of the \ion{O}{3} $\lambda$5007 \AA\ emission feature observed in NGC 1569 along the major axis, top panel, and minor axis, bottom panel.  Because the emission features are so bright, we are able to extract a single spectrum from each cross-dispersion row, giving an angular resolution of 1$\farcs09$ per pixel (17.7 pc).  Also, we achieve a velocity resolution of $\sim$27 \kms\ as measured from a comparison lamp emission line at $\lambda$ = 5014 \AA\ with a FWHM of 0.46 \AA.  This corresponds to a velocity dispersion, $\sigma$, of $\sim$12 km s$^{-1}$.

		\subsubsection{Multiple Gaussian Decomposition} \label{sec:3.2.2}
		
Figure \ref{fig:9} shows an example of a single extracted spectrum of the \ion{O}{3} $\lambda$5007 \AA\ emission feature taken from the major axis.  It is evident that there are three distinct, bright peaks over a total width of 4.1 \AA, or, $\sim$176 km s$^{-1}$. 
We disentangled these blended features with multiple Gaussian components ({\sc splot}).  
The three velocities obtained for the brightest three peaks in Figure \ref{fig:9} are -133.9, -84.2, and -31.5 km s$^{-1}$ spanning a range of over 100 km s$^{-1}$ at a distance of -13$\farcs2$ from the center of the galaxy along the major axis.  This fitting procedure was repeated for each spectrum extracted along the cross-dispersion axis.  

	\subsection{Extracting the \ion{H}{1} Bulk Velocity Field} \label{sec:3.3}

	  	\subsubsection{Tilted Ring Models} \label{sec:3.3.1}
	

We use the newly imaged LITTLE THINGS \ion{H}{1} data to determine the maximum rotation, $V_{\rm max}$, of the \ion{H}{1} gas for NGC 1569.  One common method of constructing rotation curves of disk galaxies from high-resolution \ion{H}{1} data is the tilted ring model (Begeman 1989).  The first step in applying this model is to create a two-dimensional velocity field out of a three-dimensional \ion{H}{1} data cube.  Then, from the extracted velocity map, a series of ring parameters are determined by independently fitting a series of concentric, circular rings in the plane of the disk. The ring parameters are: inclination of the galaxy disk ($\it{i}$); PA of the major axis ($\theta$) measured east from north; kinematic center ($X_{\rm pos}$, $Y_{\rm pos}$); systemic velocity ($V_{\rm sys}$); expansion velocity in the radial direction of the disk ($V_{\rm exp}$); and, finally, the rotational velocity ($V_{\rm rot}$).  The ring size and spacing between rings must be fixed and are usually determined using the beam size as the smallest ring radius and ring width.  In the tilted ring model, these rotational parameters are determined by assuming an axisymmetric thin disk with closed circular orbits and are fit independently to each of the pre-determined rings in succession.  


To determine the rotation curve of NGC 1569 we use the GIPSY\footnote{The Groningen Image Processing SYstem (GIPSY) has been developed by the Kapteyn Astronomical Institute.} task {\sc rotcur}, which applies the tilted ring model.  We input the bulk velocity field, described in the following section, along with the rotation parameters derived after eight iterations, which is how many iterations were required for the model to converge.  We used rings that were 8$\arcsec$ wide (roughly one beam width) and separated by 8$\arcsec$ so that there was no overlap between rings. Our final ring radius was 200$\arcsec$ because beyond this radius the signal-to-noise dropped drastically.  The final parameters determined from the tilted ring model are as follows: the kinematic center, ($X_{\rm pos}$ = 04:30:46.125, $Y_{\rm pos}$
 = +64:51:10.25), which is marked by the blue star on Figure \ref{fig:46} and differs by 58$\arcsec$ from the morphological center used for the stellar kinematics; $\it{i}$ = 69 $\pm$ 7; $\theta$ = 122$\arcdeg$, which is only one degree different from the PA of the major axis used for the stars; $V_{\rm sys}$ = -85 \kms, which agrees within the error of the systemic velocity derived for the stars; $V_{\rm exp}$ = 0 \kms. Table \ref{tab:1} lists these final parameter values determined from the tilted ring model.  All velocities for the \ion{H}{1} data are heliocentric velocities, the same velocity reference frame as the stellar data.

		
		\subsubsection{The Bulk Velocity Field} \label{sec:3.3.2}

NGC 1569 shows a dominance of non-circular and random motions as seen in the intensity-weighted mean velocity field in Figure \ref{fig:34}.  
The blue features across NGC 1569 show \ion{H}{1} gas with motions that are non-circular and do not follow the bulk motion of the galaxy disk.  A `typical' rotating disk generally displays a grading scale of isovelocity contours with blue-shifted (relative to the systemic motion) \ion{H}{1} on one side and red-shifted \ion{H}{1} on the opposite side. The systemic velocity of the system separates the two halves.  This is not what is observed in Figure \ref{fig:34}. Nevertheless, we were able to apply the method developed by \citet{oh08} to extract bulk rotation from chaotic systems like NGC 1569.  

An iterative double Gaussian deconvolution procedure was used to separate velocities of \ion{H}{1} clouds superimposed upon the bulk motion at a given position in the \ion{H}{1} data cube.   The deconvolution program uses only the \ion{H}{1} intensities that are above 2 $\sigma$ in the bulk velocity field.  This limits the extracted velocity fields to the inner, concentrated \ion{H}{1} disk.  The red outline in Figure \ref{fig:34} roughly identifies the region over which we extract the bulk velocity field.
 At each position within this region, if there are two distinct Gaussian peaks detected above 2 $\sigma$, then the velocities of the central wavelengths of each respective peak are compared to a model velocity field.  The velocity closest to the model velocity field is assigned to a bulk motion map while the outlying velocity is assigned to one of two non-circular motion maps.  
 The $\it{strong}$ non-circular motion map contains all velocities that are derived from Gaussian peaks that have a $\it{higher}$ intensity than the bulk motion velocity, whereas the reverse is true for the $\it{weak}$ non-circular motion map.  
Figure \ref{fig:13} shows (a) the final bulk velocity map, (b) the strong non-circular velocity map, (c) the weak non-circular velocity map, and (d) the strong non-circular motions superimposed on the weak non-circular motions map.  
The black oval in each panel of Figure \ref{fig:13} identifies an \ion{H}{1} cloud that has systemic motions significantly deviating from the bulk motion of the galaxy.  In fact, this cloud is virtually absent in the bulk velocity field (see panel (a) in Figure \ref{fig:13}).   \citet{sti02} first detected this \ion{H}{1} cloud and describe it as ``high velocity cloud 3 (HV3)''.  This cloud is discussed in more detail in Section \ref{sec:5}.  In Section \ref{sec:3.3.3}, we derive the \ion{H}{1} rotation curve using the two-dimensional bulk velocity field and the tilted ring model.

			
\section{Results} \label{sec:6}

We cannot assume that the stellar and \ion{H}{1} disks have the same geometry because there are examples of dIm systems in which the stars and gas have different kinematics.  In the case of WLM, \citet{lea09} compare the kinematics of the stars and \ion{H}{1} and discover that the stars and \ion{H}{1} are kinematically decoupled and do not possess the same geometry.  They conclude that the stars lie in a thicker disk than the \ion{H}{1}.  \citet{hun02} determined that the stars in NGC 4449 are in a face-on disk, whereas the \ion{H}{1} disk is inclined and that it exhibits rotation as well as a counter-rotation component in the center.  Therefore, we are careful here to examine the stellar kinematics and gas kinematics independently.  We begin by investigating the stars alone. 

	\subsection{Stellar Optical Spectroscopy} \label{sec:3.1.3}

Figure \ref{fig:6} shows position-velocity (P-V) graphs, where the positive positions are in the eastern direction.  The error bar in the radial direction represents the angular area of the slit over which the velocities were determined.  We used the weighted average of the minor axis velocity points to determine a systemic velocity, $V_{\rm sys}$, of -82 $\pm$ 7 km s$^{-1}$, which is denoted by the horizontal solid line in Figure \ref{fig:6}.  This value is consistent with \ion{H}{1} studies such as \citet{rea80}, $V_{\rm sys}$ = -77 $\pm$ 1 km s$^{-1}$, and \citet{deV91}, $V_{\rm sys}$ = -89 $\pm$ 5 km s$^{-1}$. But, it is inconsistent with the value obtained by \citet{sch92}, $V_{\rm sys}$ = -104 $\pm$ 4 km s$^{-1}$, who state that the \ion{H}{1} flux used to derive the velocity is confused with foreground Galactic emission.

We used the four PAs to constrain the inclination and line of nodes.  
The de-projection was derived by the following formula obtained from private communications with Vera Rubin:
\begin{equation}
V_{\rm obs} - V_{\rm sys} = \frac{V(\rm R)\ sin (i)\ cos(\eta - \theta)}{[\rm sec^2 (i) - tan^2 (i)\ cos^2(\eta - \theta)]^{1/2}}
\end{equation}
Here, $i$ is the inclination of the disk, $\eta$ is the PA of the observed point in the plane of the sky, $\theta$ is measured north through east to the receding major axis, which is the PA of the major axis (line of nodes), $V_{\rm obs}$ is the observed line-of-sight radial velocity, $V_{\rm sys}$ is the systemic velocity, and $V(\rm R)$ is the rotational velocity.  Although the velocities of the stars are quite irregular, we were still able to constrain the inclination of the disk to $i \approx$ 60$\arcdeg \pm 10\arcdeg$ and the major axis PA $\approx$ 120$\arcdeg \pm\ 8\arcdeg$ using a least squares fit.  Both of these estimates agree well with the \ion{H}{1} results (see Table \ref{tab:1}).  Furthermore, Figure \ref{fig:25} shows P-V diagrams of the stars plotted with \ion{H}{1} intensity contours.  These plots demonstrate the kinematic similarity between the stars and \ion{H}{1}, as the stars follow the \ion{H}{1} in all four PAs. 
The stellar velocity dispersions are shown in Figure \ref{fig:20} for each of the four PAs.  The average of the velocity dispersion, $<\sigma_{\rm obs}>$, weighted by the uncertainty of each point, for all four PAs is 21 $\pm$ 4 km s$^{-1}$. 

 In order to disentangle $\sigma_{\rm R}$, $\sigma_{\rm z}$, and $\sigma_{\rm \phi}$ from $\sigma_{\rm obs}$, 
we use the following relation from \citet{ger97}:
\begin{equation}\label{equ:3}
\sigma_{\rm obs}^2 = [\sigma_{\rm R}^2\ \rm sin^2(\eta - \theta)  + \sigma_{\rm \phi}^2\ cos^2(\eta - \theta )] sin^2i + \sigma_{\rm z}^2\ cos^2i
\end{equation}
Here, $\eta$ is the intrinsic angle within the plane of the galaxy, $\theta$ is the receding major axis PA and $i$ is the inclination of disk.  Since the \ion{H}{1} and stars are well matched (see Section \ref{sec:4.1.2}), from here on we use the better determined inclination of $i$ = 69$\arcdeg$ $\pm\ 7\arcdeg$ and major axis PA of 122$\fdg5$ from the \ion{H}{1} kinematics.
Since we have only a few points along each intermediate PA, i.e., $\pm$ 45$\arcdeg$ from the major axis, we focus on the major axis, where $\eta = \theta$ and thus, $\sigma_{\rm R}$ does not contribute to $\sigma_{\rm obs}$.   Following Swaters (1999), we make the assumption that the stellar velocity dispersion in the stars follows the Milky Way properties observed for stars near the Sun. Therefore, $\sigma_{\rm obs}$ = $\sigma_{\rm z}$ for the major axis.  

Swaters (1999) found a relation between integrated $M_{\rm B}$ of a galaxy and the central vertical velocity dispersion:
\begin{equation}
{\rm log}\ \sigma_z = -0.15M_{\rm B} - 1.27
\end{equation}
This relation predicts a $\sigma_z$ of 21 km s$^{-1}$ for NGC 1569, which is equal to our observed value of 21 $\pm$ 4 km s$^{-1}$.  NGC 1569 agrees well with the overall trend observed in spiral galaxies, i.e., lower luminosity systems have lower velocity dispersions.

	\subsection{Ionized Gas Spectroscopy}  \label{sec:4.1.2}
	
Figure \ref{fig:10} shows the decomposed line-of-sight radial velocities of the ionized gas for the major axis and has the stellar and \ion{H}{1} velocities over plotted for comparison.  
The \ion{H}{1} velocities in this figure come from a P-V slice through the major axis of the \ion{H}{1} data cube with a sampling width of 8$\arcsec$, equal to the FWHM of the \ion{H}{1} beam.  There was no deconvolution done to produce these \ion{H}{1} velocities; they are simply the observed velocities along the major axis position angle. The velocities of the \ion{O}{3} ionized gas match the velocities of H$\alpha$ along the major axis observed by \citet[see their Figure 3]{tom94}.

In Figure \ref{fig:10}, we observe two forms of ionized gas: high velocity gas and low velocity gas.  The low velocity ionized gas has velocities that match the \ion{H}{1} and stellar velocities and is likely associated with HII regions, or the remnants thereof, leftover from recent star formation.  The high velocity ionized gas at a radius of -20$\arcsec$ has the signature of an expanding shell as identified by the solid red outline in Figure \ref{fig:10}. This expanding shell has an expansion velocity of $\sim$70 km s$^{-1}$.  The stellar velocities over plotted in Figure \ref{fig:10} perfectly match the velocities of the low velocity gas ($\sim\ +$10 km s$^{-1}$) at the center of the expanding gas shell (R $\sim$ $-$20$\arcsec$).  These low velocities of the ionized gas also coincide with the bulk motion of the \ion{H}{1} gas.  Curiously, this agreement extends to the peculiar dip to lower velocities at the western edge of the red oval outline.  This agreement in velocity (stars and \ion{O}{3}) shown by the black arrow in Figure \ref{fig:10}, are presumably from HII regions connected with particularly bright stars.  Within this expanding shell, there is only one \ion{H}{1} velocity point.  The absence of \ion{H}{1} in this region is due to an \ion{H}{1} hole blown out by SSC A \citep{isr90}. SSCs A and B reside at radii -22$\arcsec$ and -12$\arcsec$, respectively.   
Due to the lack of \ion{H}{1} and the matching stellar and low velocity ionized gas in the expanding shell region, even including peculiar motions, we conclude that our stellar spectra are dominated by a young, luminous stellar population that are ionizing nearby gas left over from the star formation process.  At the same time these young, luminous stars, mostly from SSC A, created the expanding shell from winds and supernova explosions during the most recent star formation episode. This expanding shell was also observed and discussed by \citet{wes07a, wes07b}, who also observed both the low and high velocity ionized gas.

The high velocity ionized gas seen everywhere else is likely the result from outflow from the recent starburst activity \citep{wes07a, wes07b}.  We also see evidence of galactic wind at radii greater than $+$0$\arcsec$.  Because of the exponential drop in stellar light, the integrated emission from the stars was too faint to extract stellar kinematic information in our optical spectra beyond a radius of $\pm$30$\arcsec$.  Thus, no stellar velocity information is known in the region of the high velocity ionized gas at radii greater than $+$30$\arcsec$.


The low velocity ionized gas centered at R = $+$15$\arcsec$ in Figure \ref{fig:10} also shows matching stellar and gas velocities.  This region, however, has more \ion{H}{1} than the expanding shell centered at $-$20$\arcsec$. It is not as obvious here as for the expanding shell centered at R = $-$20$\arcsec$, but it is possible that the stellar spectra around R = $+$15$\arcsec$ are also dominated by young, luminous stars because they, too, follow the velocities of the low velocity ionized gas and \ion{H}{1}.  

Studies of HII regions of NGC 1569 \citep{gre96, hun93, wal91, got90, you84}
show that there is star formation in the region at R = $+$50$\arcsec$, but it is not as intense as the western side of the galaxy where the SSCs reside.
Again, we note that the velocities of the ionized HII regions (low velocity gas) and \ion{H}{1} agree well in this area.

	\subsection{\ion{H}{1} rotation curve} \label{sec:3.3.3}
	
Figure \ref{fig:15} shows the observed rotation curve.  We find a maximum rotational velocity, $V_{\rm rot}$, of 33 $\pm$ 10 km s$^{-1}$, which was calculated from the average of the last nine data points in Figure \ref{fig:15}.  The uncertainty is dominated by the rapid drop in signal-to-noise in the bulk velocity field beyond a radius of 2.2 kpc.  Within the errors, Figure \ref{fig:15} shows a rising, solid body-type rotation curve, which at best reaches the usual flat part at radii $>$ 2.2 kpc. 
	
Figure \ref{fig:59} shows a velocity dispersion map of the bulk (main) velocity component only.  The average \ion{H}{1} velocity dispersion is high, $<\sigma_{\rm HI}>$ = 10 km s$^{-1}$, compared to the observed rotation speed, $V_{\rm rot}$ = 33 $\pm$ 10 km s$^{-1}$.
 Because of the high \ion{H}{1} velocity dispersions in NGC 1569, we correct the observed rotation velocities for asymmetric drift in the same manner as \citet{oh11}, and derive the curve shown by the filled circles in Figure \ref{fig:50}.  
 The maximum rotation velocity determined from \ion{H}{1}, increased from the observed $V_{\rm rot}$ = 33 km s$^{-1}$ to $V_{\rm max}$ = 50 km s$^{-1}$ after the asymmetric drift correction was applied. 

\section{Discussion} \label{sec:5}	 
	
	\subsection{The Inner Kinematic Structure of NGC 1569} \label{sec:5.1}
	

The upper left panel in Figure \ref{fig:25} displays the observed radial velocity of gas (contours) and integrated stars (points) as a function of position along the slit that passes through the major axis.  This P-V diagram shows ``forbidden'' \ion{H}{1} motion in the upper right quadrant between -40 and -70 km s$^{-1}$ at around a distance of -0$\farcm6$ from the galaxy center, as identified by the black oval.  We define ``forbidden'' \ion{H}{1} as motion that does not follow the ordered rotation.  Along the major axis in Figure \ref{fig:25}, we show a black curve that traces the average velocities of the \ion{H}{1} contours from 0$\arcmin$ to 2$\arcmin$, which is the region of the diagram that follows the ordered motion of the \ion{H}{1} gas.  This black curve is then reflected through the origin to show what the expected ordered motion should be from -2$\arcmin$ to 0$\arcmin$. 
At some positions between distances of 0$\arcmin$ and -0$\farcm5$ from the galaxy center, the stellar velocities follow the velocity trend of the forbidden motion,  whereas at other positions around -0$\farcm4$, the stellar velocities follow the ordered rotation of the \ion{H}{1} disk.  

\begin{figure}[H]
\centering
\includegraphics[scale=.3]{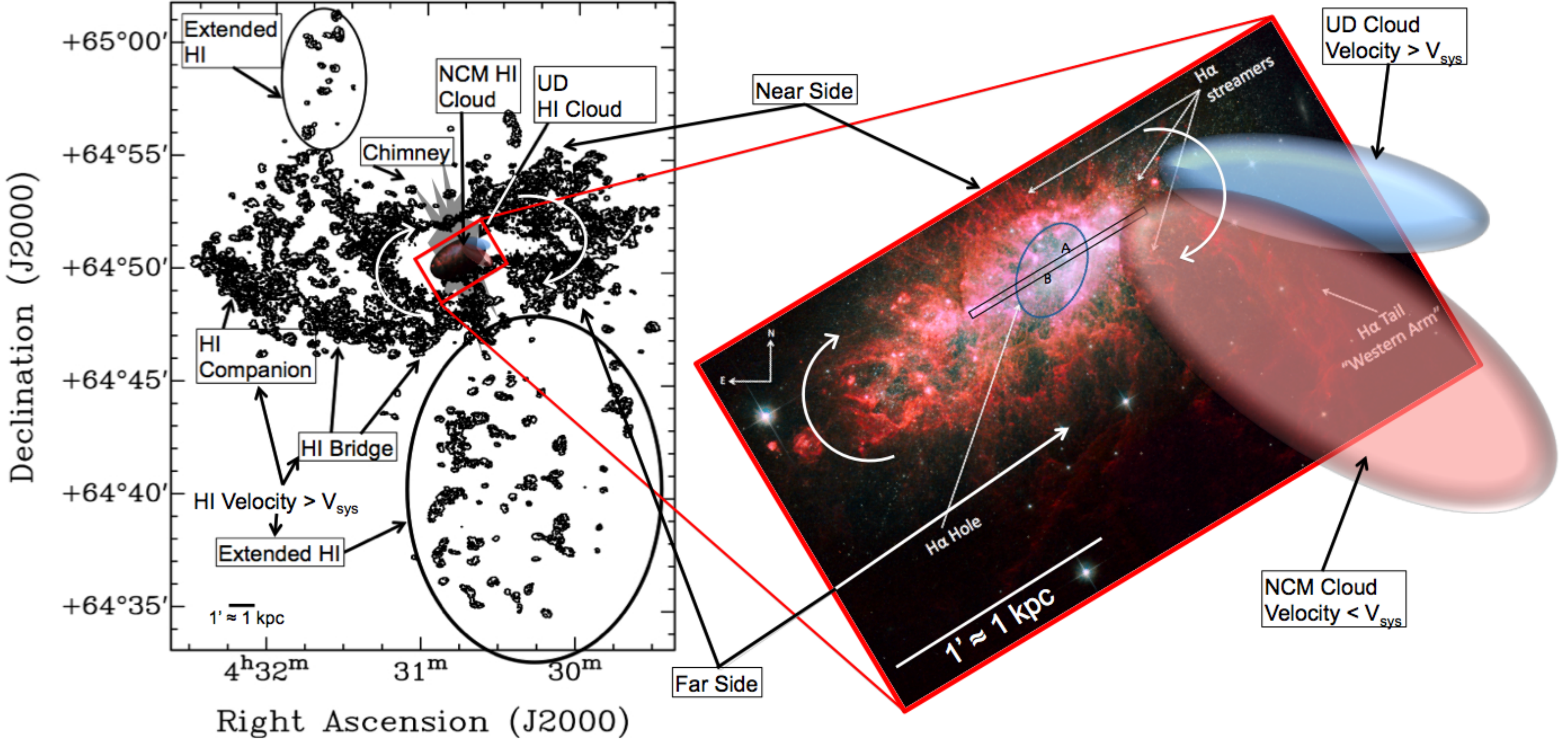}
\caption{Outline of all major gas features in NGC 1569. The black contours in the left image are from the integrated \ion{H}{1} data and the center of the map is left blank so that the scale of the {\it HST} image and NCM \ion{H}{1} cloud and UD \ion{H}{1} cloud can be seen.  The extended \ion{H}{1} to the northeast and south of NGC 1569 were imaged for the first time in this work.  The right image is an {\it HST} composite image showing the \ha\ features along with the UD and NCM \ion{H}{1} clouds.  SSCs A and B are identified in the right {\it HST} image as well as the 
position, location and 
size of the optical slit used to derive the stellar kinematics. The white curved arrows in both images show the rotation of the disk.  The near and far sides of the disk have been labeled as well as the relative velocities of the \ion{H}{1} clouds and features. Credit: NASA, ESA, the Hubble Heritage Team (STScI/AURA), and A. Aloisi (STScI/ESA)}
\label{fig:cartoon} 
\end{figure}

\begin{figure}[H]
\centering
\includegraphics[scale=.5]{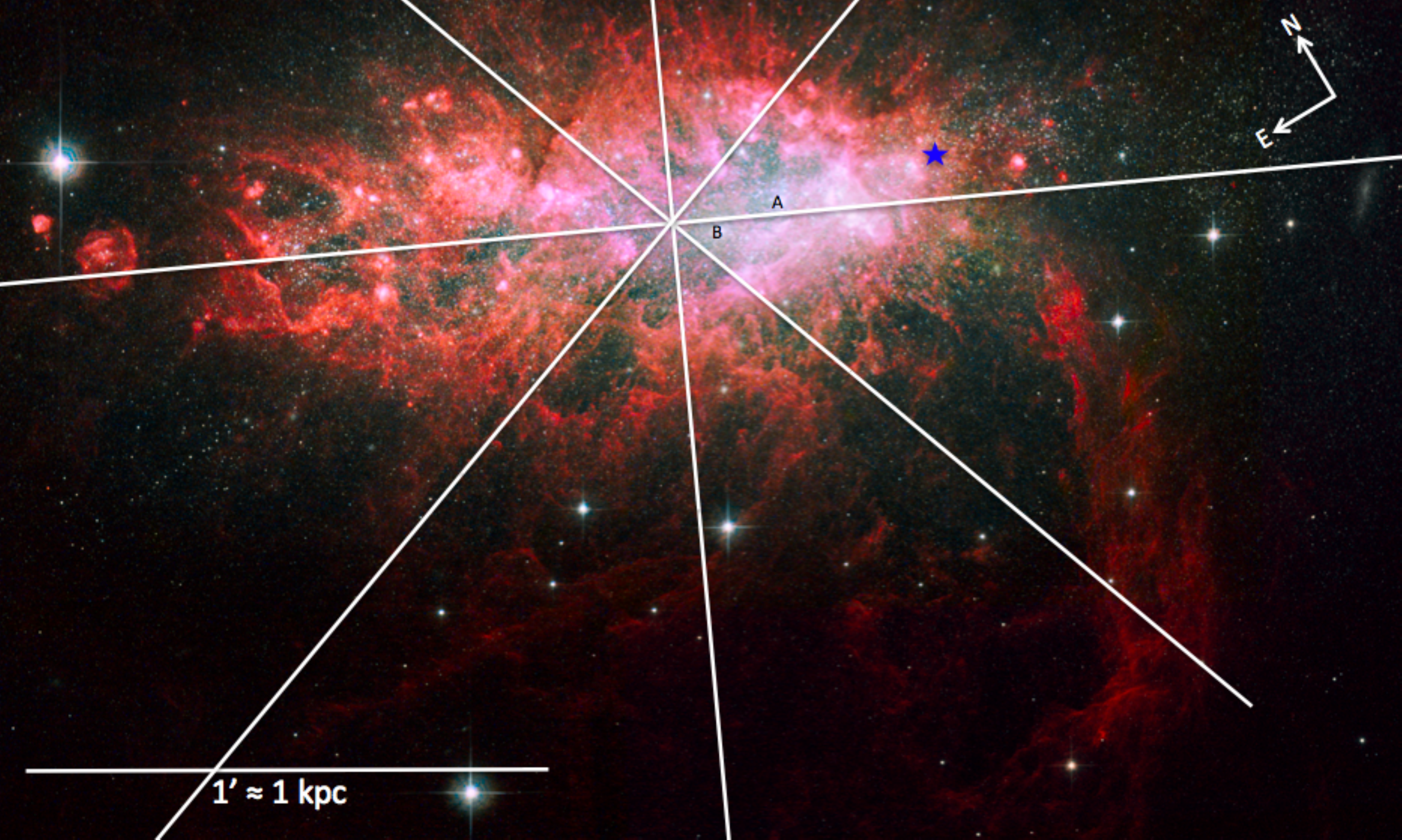}
\caption{
{\it HST} image from Figure \ref{fig:cartoon} showing the slit positions and the location of SSCs A and B. The bar at the bottom left in both panels shows $\sim$ 1 kpc in distance and the blue star marks the kinematic center determined from the \ion{H}{1} tilted ring model.}
\label{fig:46}
\end{figure}

\begin{figure}[H]
\centering
\includegraphics[angle=90,scale=.4]{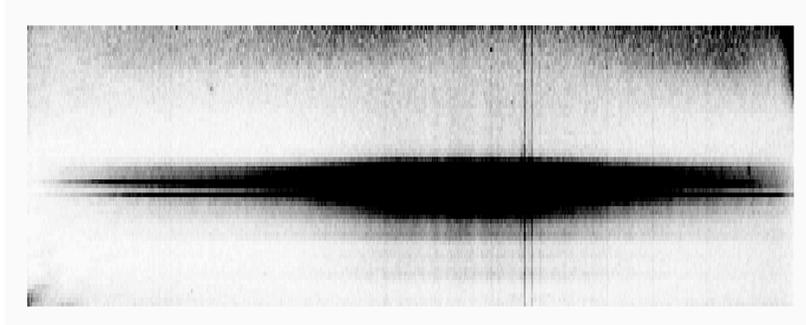}
\hfil
\includegraphics[angle=90,scale=.4]{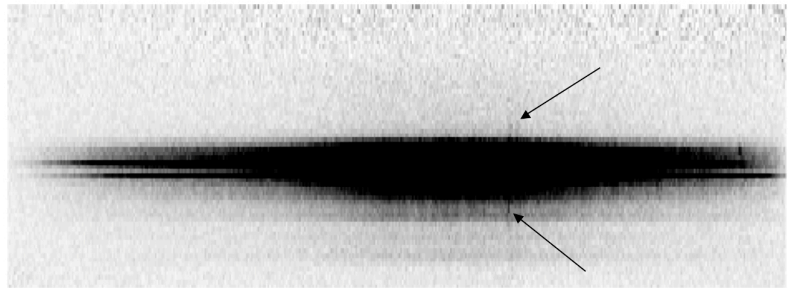}
\caption{\underline{Top}: Five 1800s spectra co-added to create a two-dimensional galaxy spectrum of the major axis of NGC 1569 \emph{before} sky subtraction.  \underline{Bottom}:  Same as top only \emph{after} sky subtraction.  For both panels, wavelength is along the horizontal axis ($\sim$238 \AA\ total coverage) and the vertical axis is the cross-dispersion direction along the slit ($\sim$3$\arcmin$).  The right panel shows residual sky emission lines, indicated by the arrows.  Only the bright region in the spatial center was used in the cross-correlation process.}
\label{fig:2}
\end{figure}

\begin{figure}[H]
\centering
\includegraphics[angle=90,scale=.4]{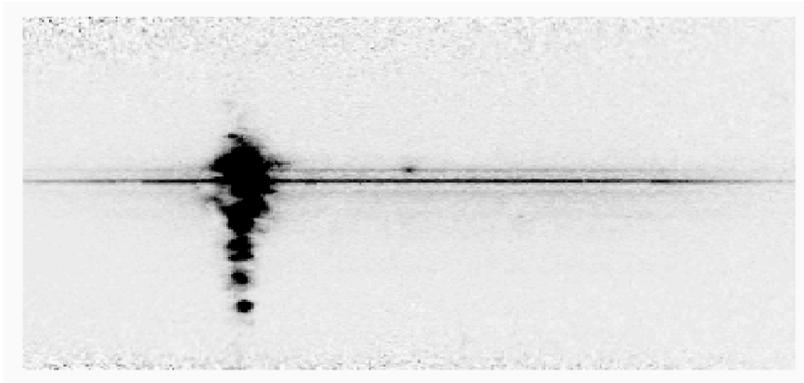}
\hfil
\includegraphics[angle=90,scale=.4]{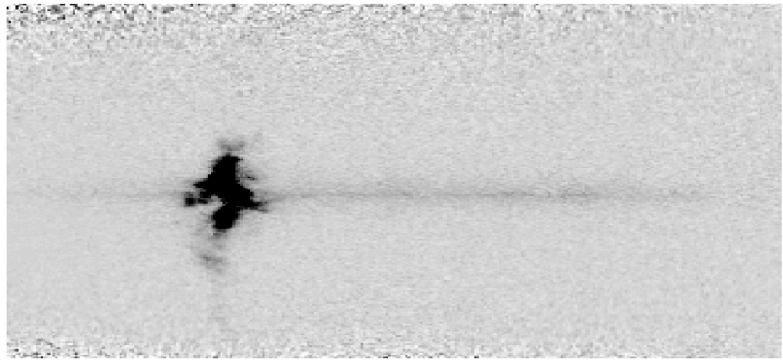}
\caption{\underline{Top}: Co-added, sky subtracted, two-dimensional spectrum of the \ion{O}{3} $\lambda$5007 \AA\ emission feature for the major axis of NGC 1569. \underline{Bottom}: Same as left, only for the minor axis. Wavelength is along the horizontal axis ($\sim$41 \AA\ total coverage) and the vertical axis is the cross-dispersion direction of the slit ($\sim$3$\arcmin$).}
\label{fig:3}
\end{figure}

\begin{figure}[H]
\centering
\includegraphics[scale=0.6]{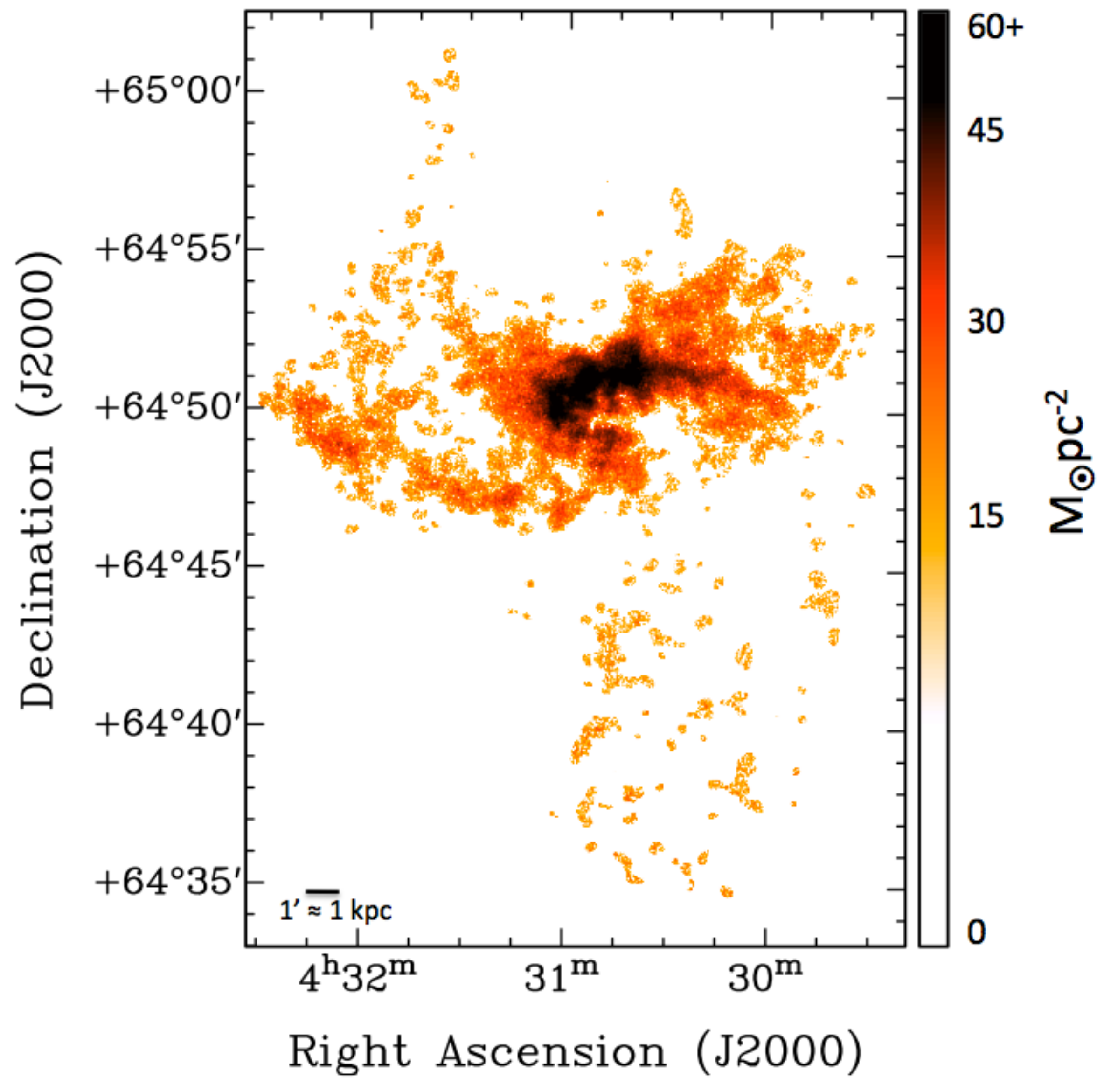}
\caption{
Integrated \ion{H}{1} flux of NGC 1569.  The multi-scale cleaning algorithm implemented by LITTLE THINGS was able to image tenuous emission to the south and northeast for the first time.}
\label{fig:47}
\end{figure}

Could the gas and stars be showing evidence of counter-rotation?  If this were the case, then we would expect \ion{H}{1} contours in the lower left quadrant of all the panels of Figure \ref{fig:25}, which we do not observe.  Therefore, 
we identify the forbidden \ion{H}{1} motion as 
the northeastern tip of the NCM cloud identified in Figure \ref{fig:cartoon} and by the black oval in Figure \ref{fig:13}.  The major axis, PA = 122$\arcdeg$ (shown in Figure \ref{fig:cartoon} and by the black line throughout Figure \ref{fig:13}), slices through the tip of the NCM cloud. The velocities of this cloud in the strong non-circular motion map, panel (b) of Figure \ref{fig:13}, match the velocities of the contours of the forbidden \ion{H}{1} in Figure \ref{fig:25}.  The NCM cloud has similar velocity structure and position as HV3 in \citet{sti02}.  We conclude that the stars and gas are not counter-rotating but, there is evidence of a kinematic disturbance.

\begin{figure}[H]
\centering
\includegraphics[scale=0.4]{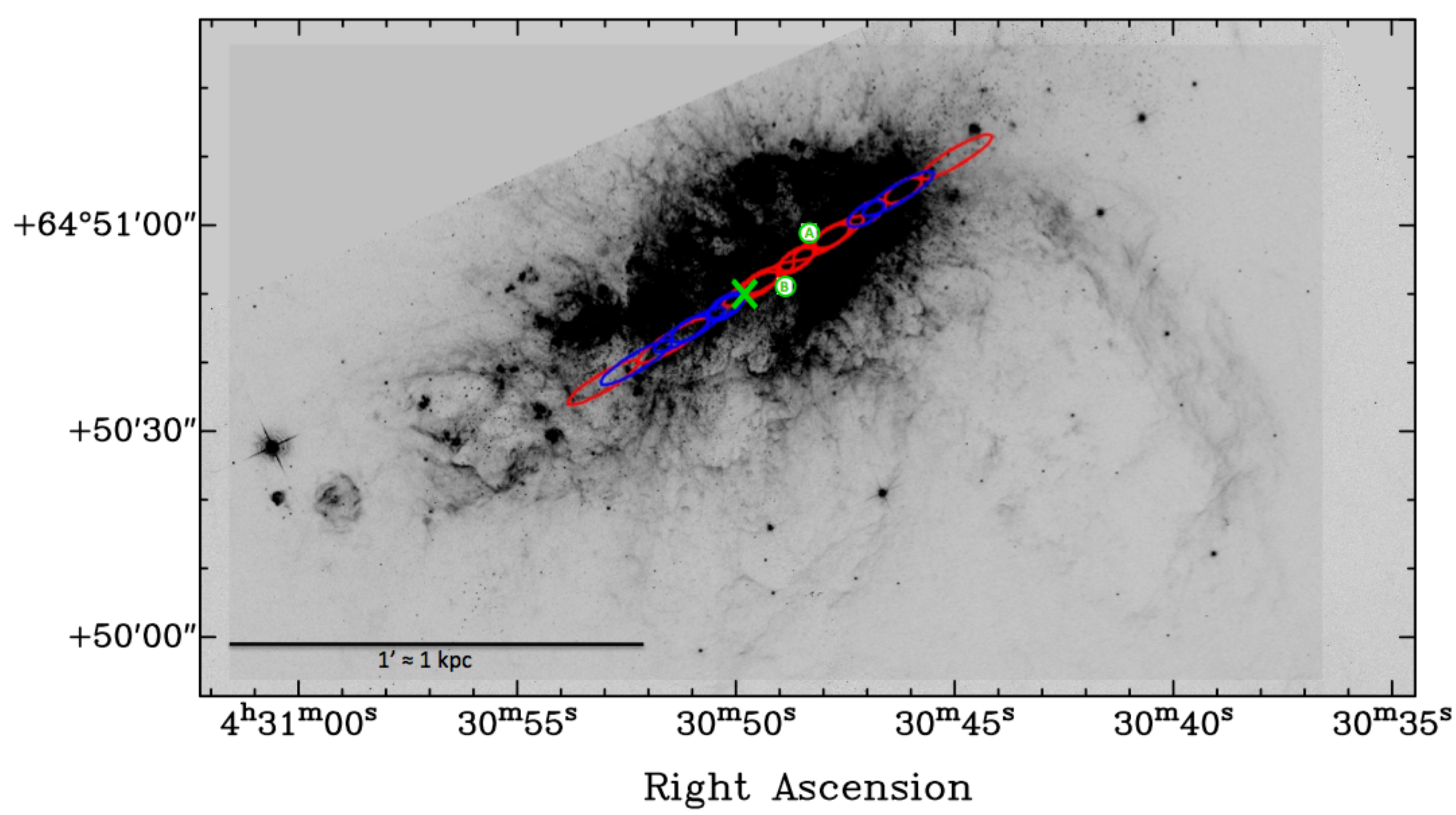}
\caption{
Same {\it HST} image as in Figure \ref{fig:cartoon} only shown in greyscale with the major axis slit superposed.  The ovals are the size and position of the extracted stellar spectra.  The red ovals correspond to red-shifted velocities relative to the systemic velocity whereas blue ovals correspond to blue-shifted velocities. The green `X' marks the center of the slit and the galaxy. SSC A and B are marked by the green circles.  Credit: NASA, ESA, the Hubble Heritage Team (STScI/AURA), and A. Aloisi (STScI/ESA)}
\label{fig:29}
\end{figure}

\begin{figure}[H]
\centering
\includegraphics[scale=.2]{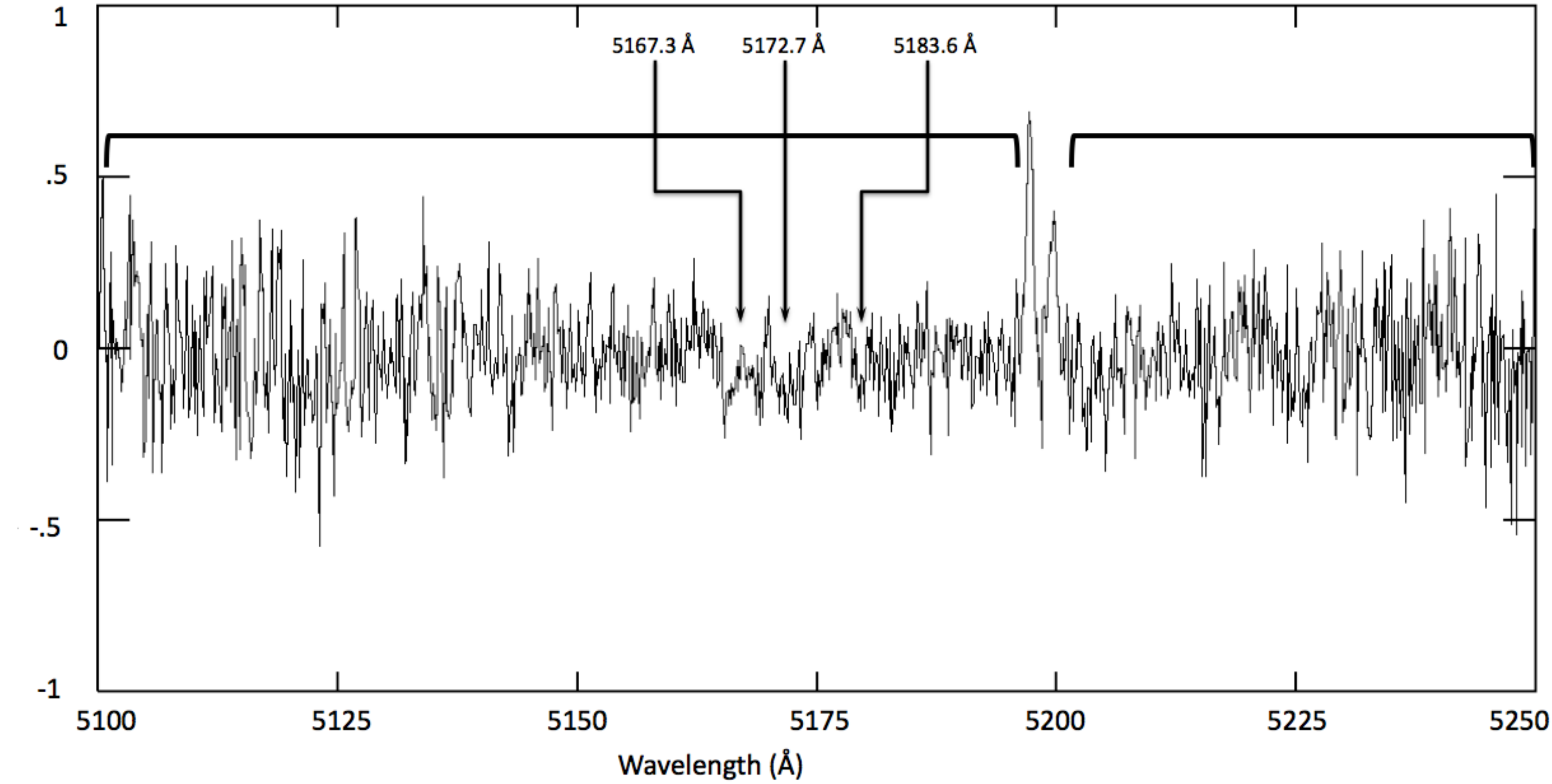}
\hfil
\includegraphics[scale=.2]{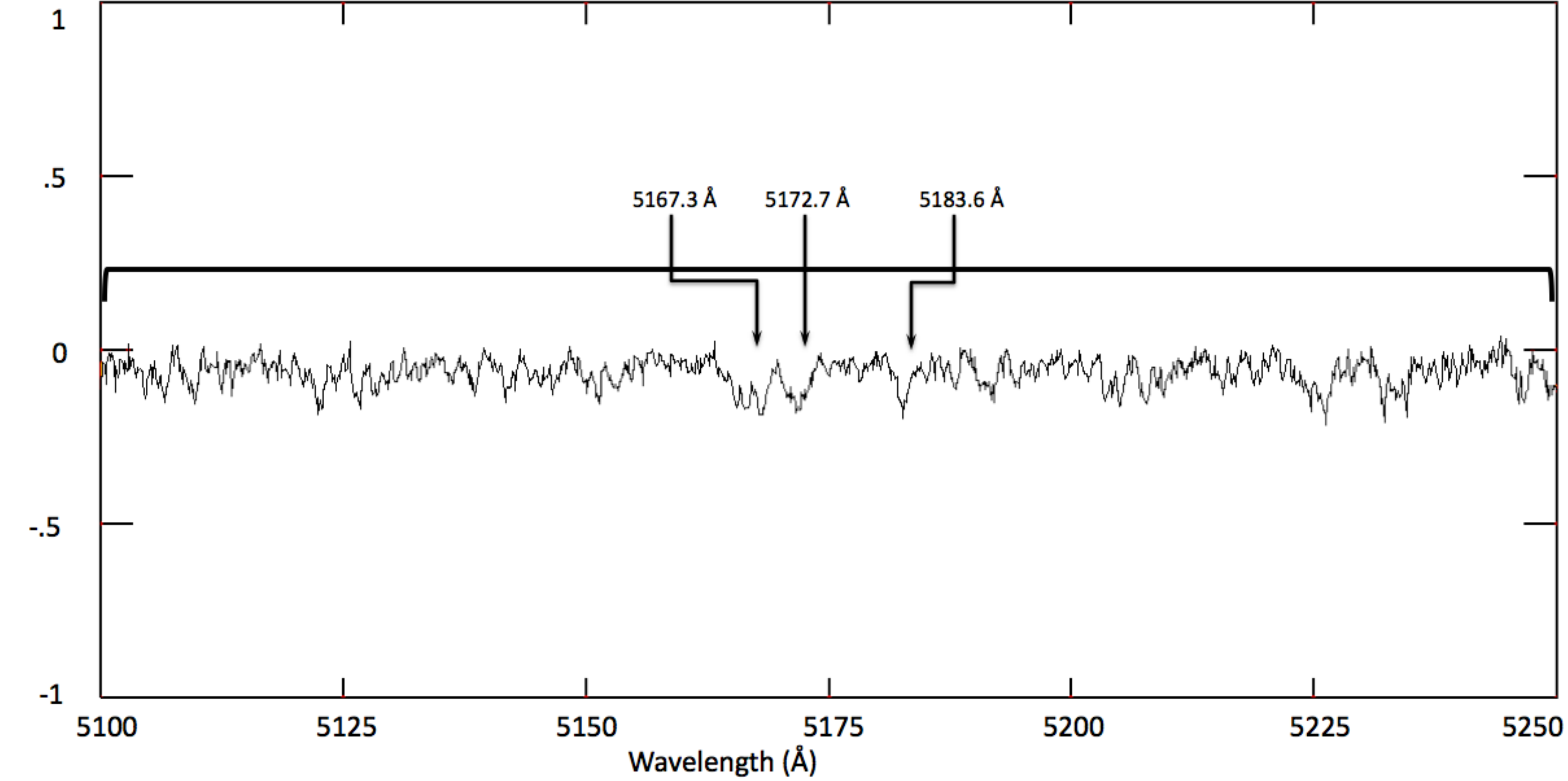}
\hfil
\includegraphics[scale=.21]{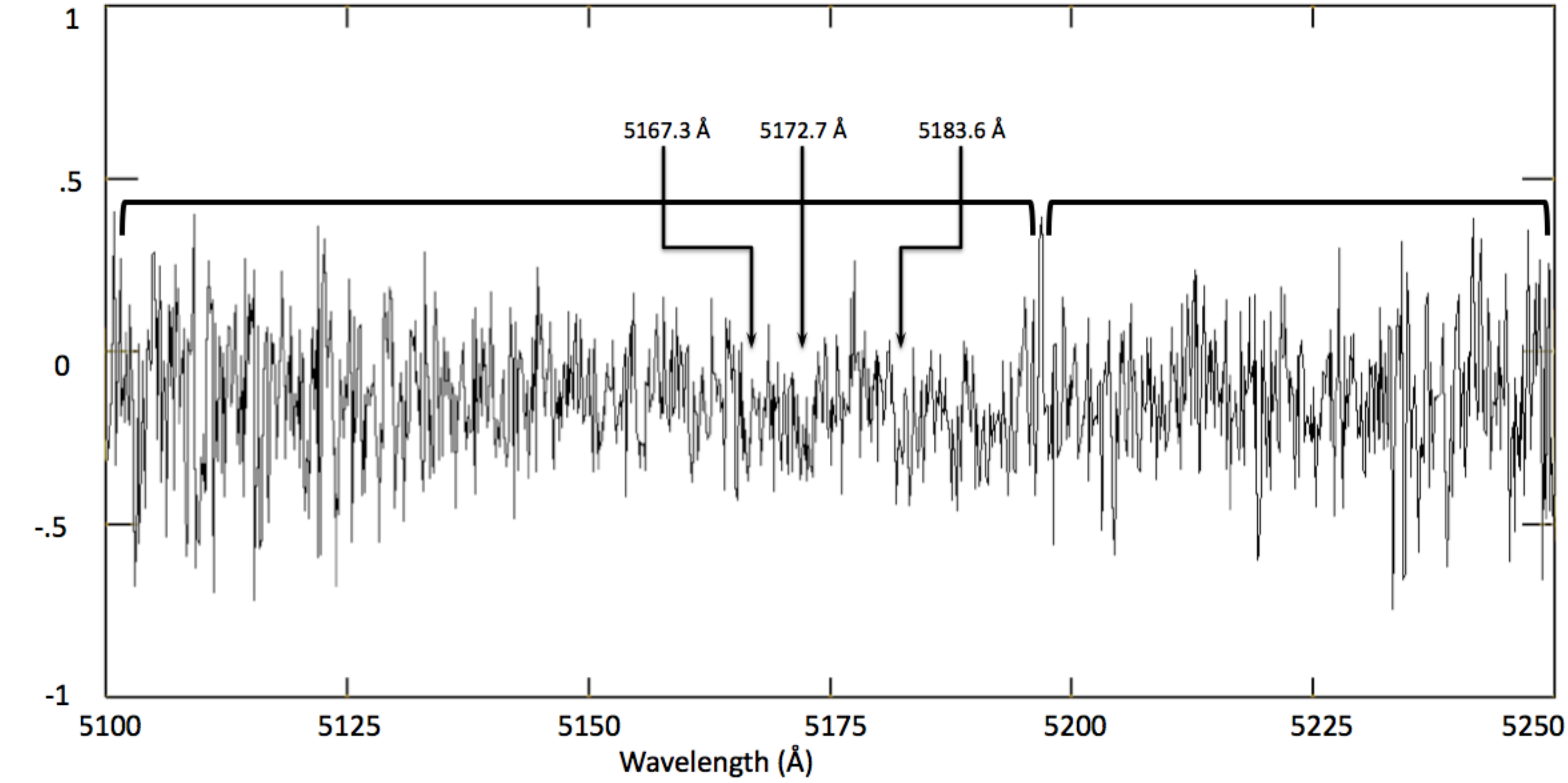}
\hfil
\includegraphics[scale=.21]{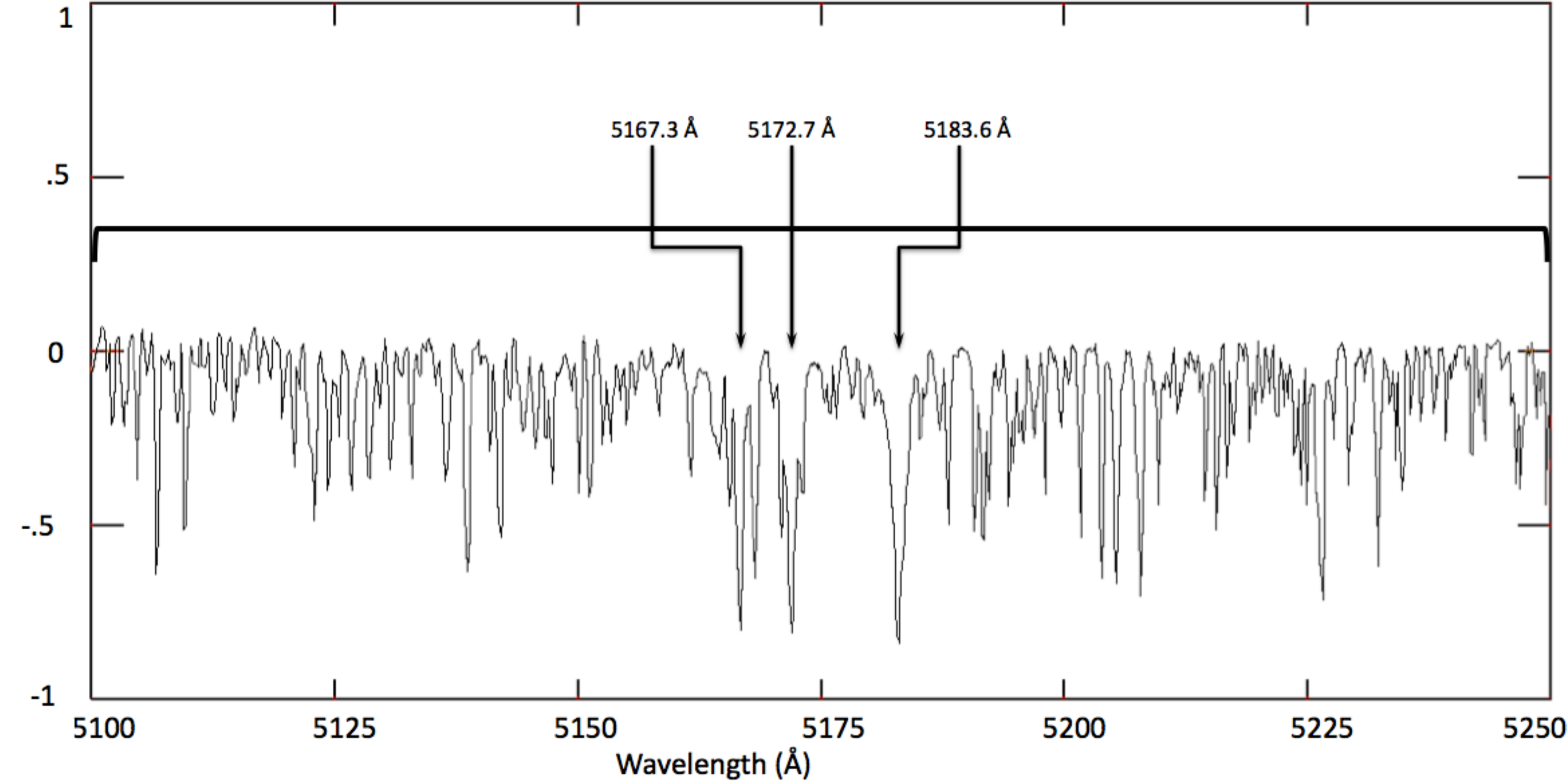}
\caption{\small Example of three, one-dimensional galaxy spectra (first three panels) and one template spectrum (last panel) for comparison. 
The galaxy spectra, observed along the major axis of NGC 1569, are representative of the range of signal-to-noise (S/N) and span the spatial area of the slit; the top panel has a S/N of $\sim$ 7 at a radius of +18$\arcsec$(east) from the center, the second panel has S/N of $\sim$ 24 at a radius of -4$\arcsec$(west) from the center, the third panel has a S/N of $\sim$ 6 at a radius of -34$\arcsec$(west) from center.  The stellar template spectra has a S/N of $\sim$ 100, which was typical throughout both observing runs.
We used the regions shown by the horizontal brackets in the cross-correlation method (CCM). The Mg Ib absorption features are marked in each panel.}
\label{fig:4}
\end{figure}

\begin{figure}[H]
\includegraphics[angle=90,scale=.5]{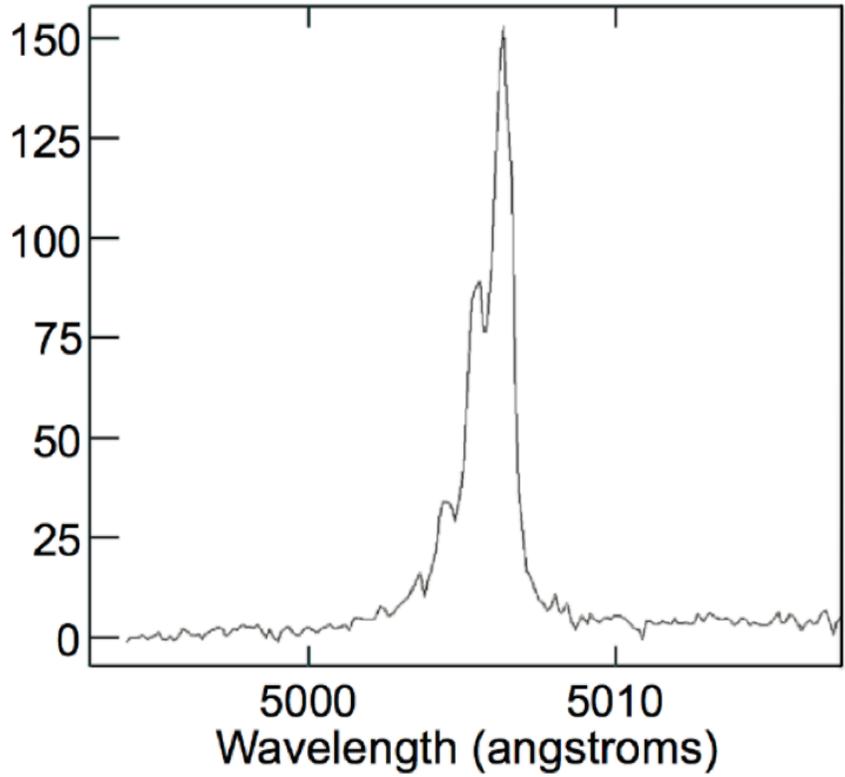}
\caption{A single extracted spectrum of the \ion{O}{3} $\lambda$5007 \AA\ emission feature at a radius of -13$\arcsec$ from the center of NGC 1569 along the major axis. Three bright Gaussian peaks are seen. The total width of the blended emission feature at zero intensity is $\sim$4.1 \AA, or, $\sim$176 km s$^{-1}$.}
\label{fig:9}
\end{figure}

\begin{figure}[H]
\includegraphics[scale=.5]{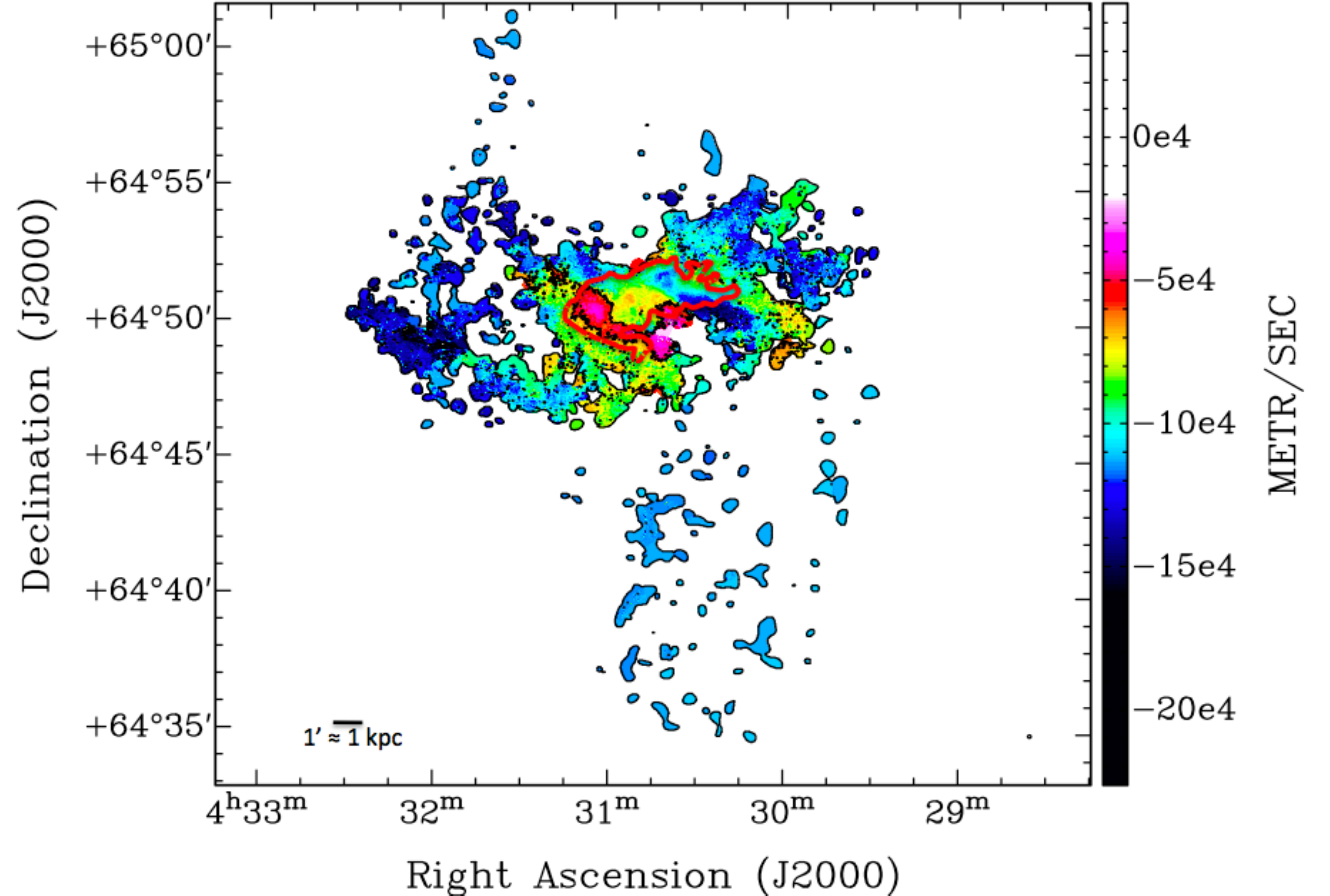}
\caption{
Intensity-weighted \ion{H}{1} velocity field of NGC 1569.  The red outline identifies the region over which the bulk motion extraction procedure was performed.  The black line marks a $\sim$1 kpc distance scale.  The tiny dot in the lower right corner near (04:29:35, +64:35:00) shows the resolution.}
\label{fig:34}
\end{figure}

\begin{figure}[H]
\centering
\includegraphics[scale=.4]{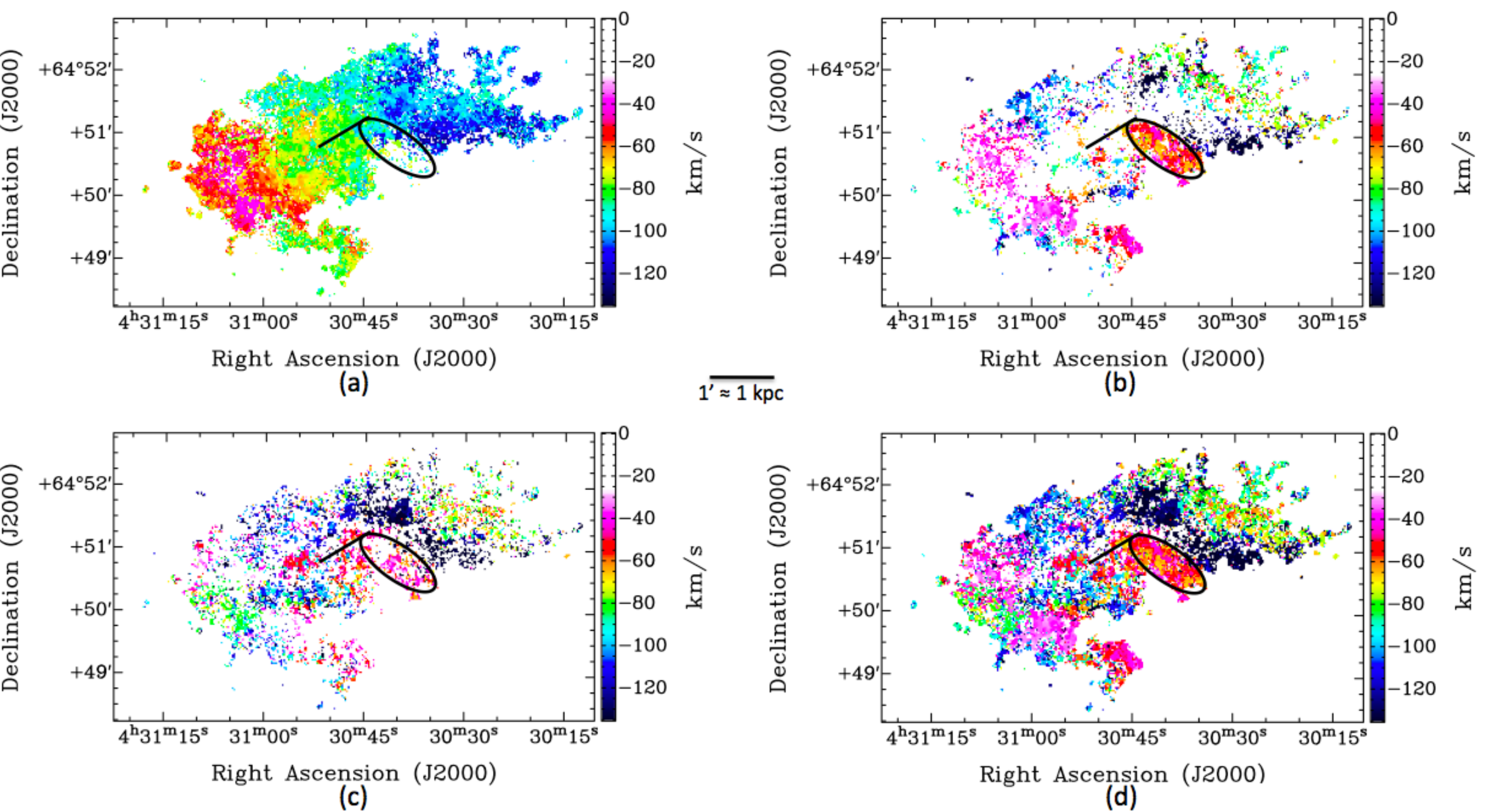}
\caption{Results of the kinematic deconvolution procedure of \citet{oh08}: (a) Bulk rotation map. (b) Strong non-circular motion map. (c) Weak non-circular motion map. (d) Combined strong $+$ weak motion map.  The black oval identifies a strong non-circular motion (NCM) \ion{H}{1} cloud at -50 \kms, clearly redshifted with respect to the underlying (bulk) motion.  The black line in the center of each velocity field identifies the position and length of the slit along the major axis used in the stellar spectroscopy.  The black line in the center of the figure shows approximately 1 kpc distance.}
\label{fig:13}
\end{figure}

\begin{figure}[H]
\includegraphics[angle=90,scale=.5]{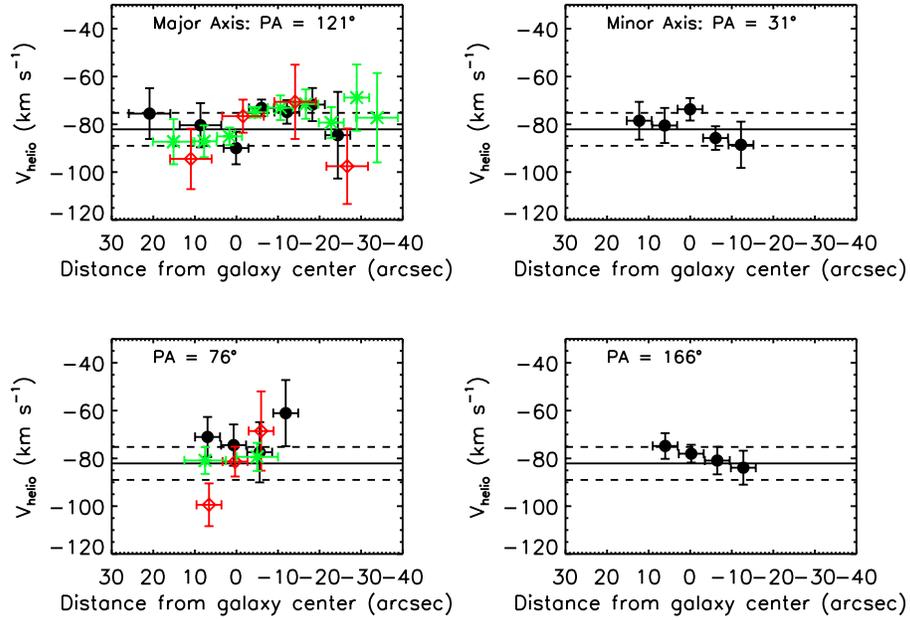}
\caption{Observed heliocentric velocities in the stars along four PAs.  Red diamonds, green stars, and black filled circles represent independent points derived from different observing nights.  The dashed lines represent the $\pm$6.9 km s$^{-1}$ uncertainty in $V_{\rm sys}$.  East are positive distances from the galaxy center.  Error bars along the x-axis show the angular area summed along the slit.}
\label{fig:6}
\end{figure}

\begin{figure}[H]
\includegraphics[angle=90,scale=.5]{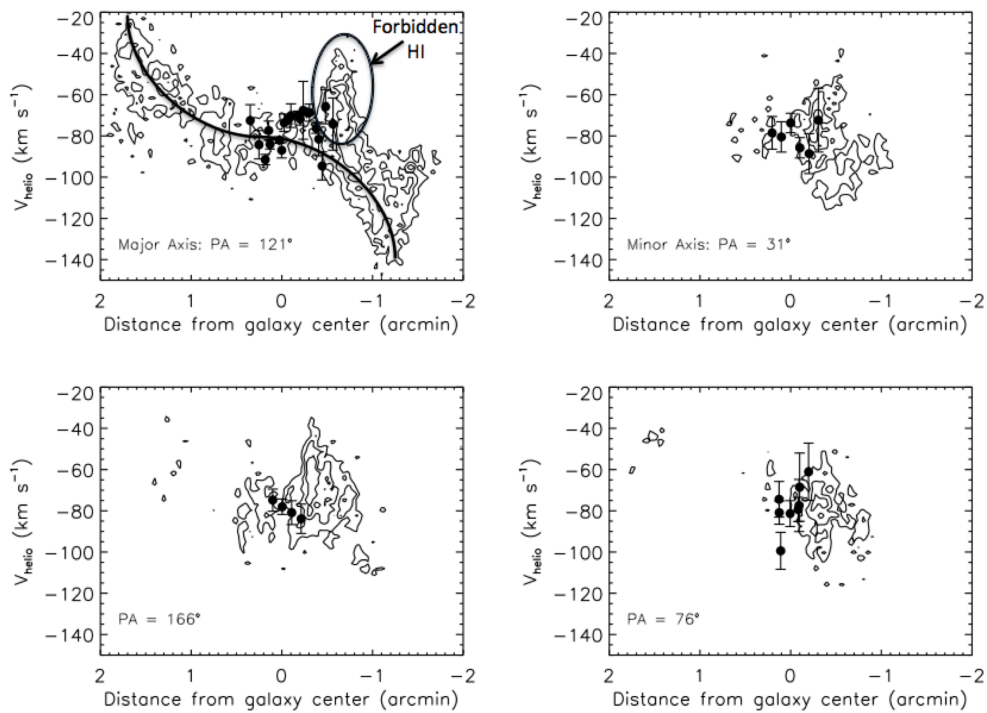}
\caption{Position-velocity diagrams taken along the four optical slit PAs showing \ion{H}{1} contours with stellar data over plotted. Positive distance corresponds with eastward direction.  The solid line in the upper left plot, major axis, shows the ordered rotation of the \ion{H}{1}.  The oval in the upper left plot identifies the ``forbidden'' \ion{H}{1} motion. The stars and the \ion{H}{1} follow each other closely at all four PAs.}
\label{fig:25}
\end{figure}

\begin{figure}[H]
\includegraphics[angle=90,scale=.5]{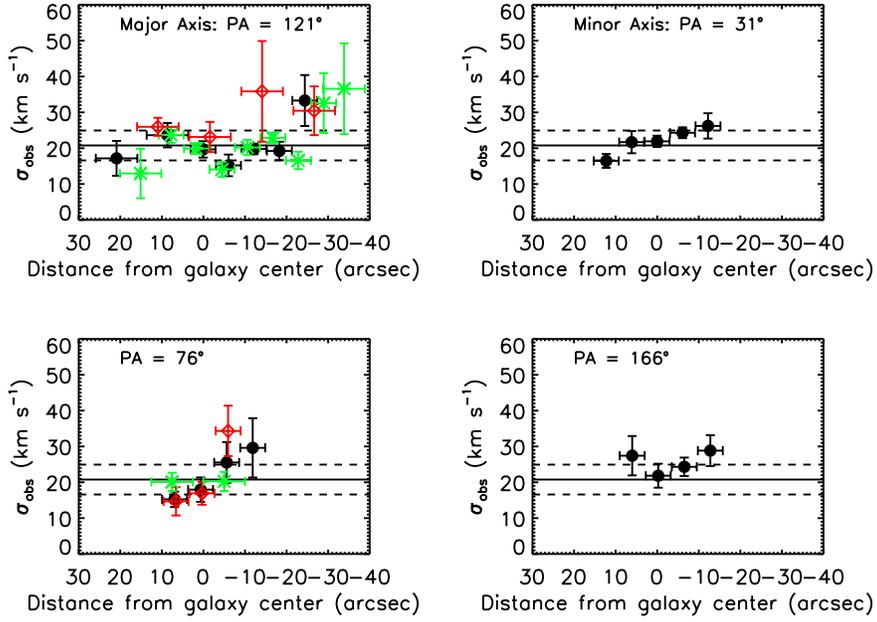}
\caption{Stellar velocity dispersions observed along the four optical slit PAs.  Black filled circles, red diamonds, and green stars represent independent points derived from different observing nights. The solid line shows the average velocity dispersion, $<\sigma_{\rm z}>$ = 21, and the dotted lines indicate the uncertainty, $\pm$ 4.2 km s$^{-1}$.  Error bars along the x-axis show the angular area summed along the slit. East are positive distances from the galaxy center.}
\label{fig:20} 
\end{figure}

\begin{figure}[H]
\includegraphics[angle=90,scale=.5]{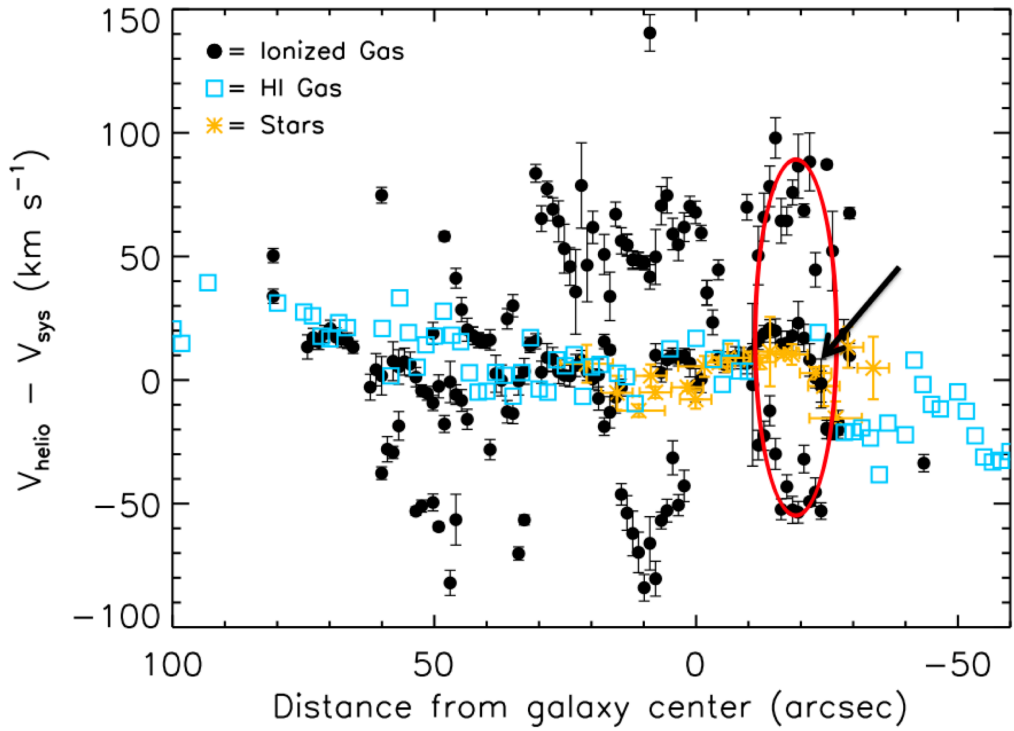}
\caption{Position-velocity plot of the ionized gas (black filled circles), \ion{H}{1} gas (blue boxes), and stars (yellow stars) along the major axis.  The solid red outlines an expanding gas shell.  Note the close correspondence between the stars and ionized gas, including peculiar motions, in the western-most oval as indicated by the black arrow.  East are positive distances from the galaxy center.}
\label{fig:10}
\end{figure}

\begin{figure}[H]
\includegraphics[angle=90,scale=.5]{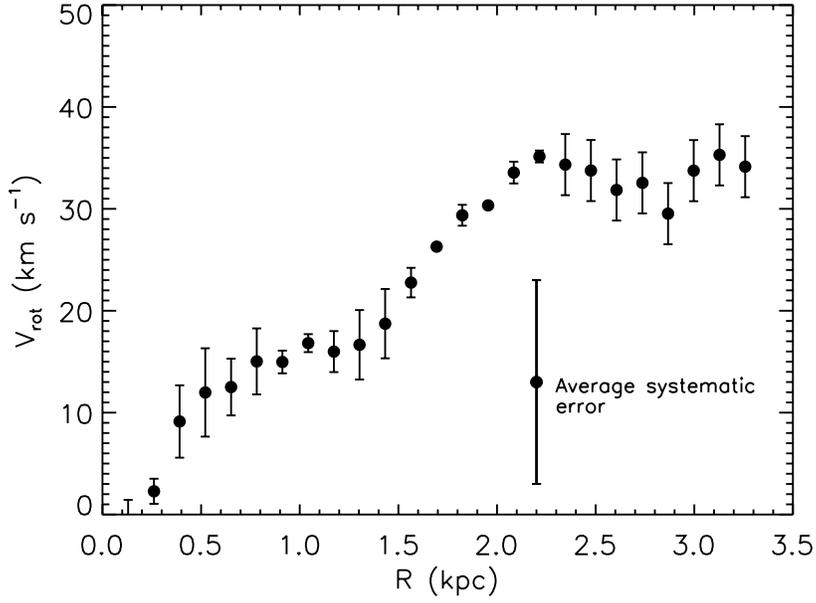}
\caption{Measured rotation curve of the \ion{H}{1} gas determined from the bulk motion map using a tilted ring model.  A $V_{\rm rot}$ of 33 km s$^{-1}$ is determined from averaging over the region where the curve is flat from $\sim$2.0 - 3.2 kpc.  The errors on the rotation curve points are the standard deviations calculated by assuming a solid body rotation out to 2.2 kpc.  The average systematic error, determined from all observational and computational uncertainty, is shown for reference.}
\label{fig:15}
\end{figure}

\begin{figure}[H]
\includegraphics[scale=.5]{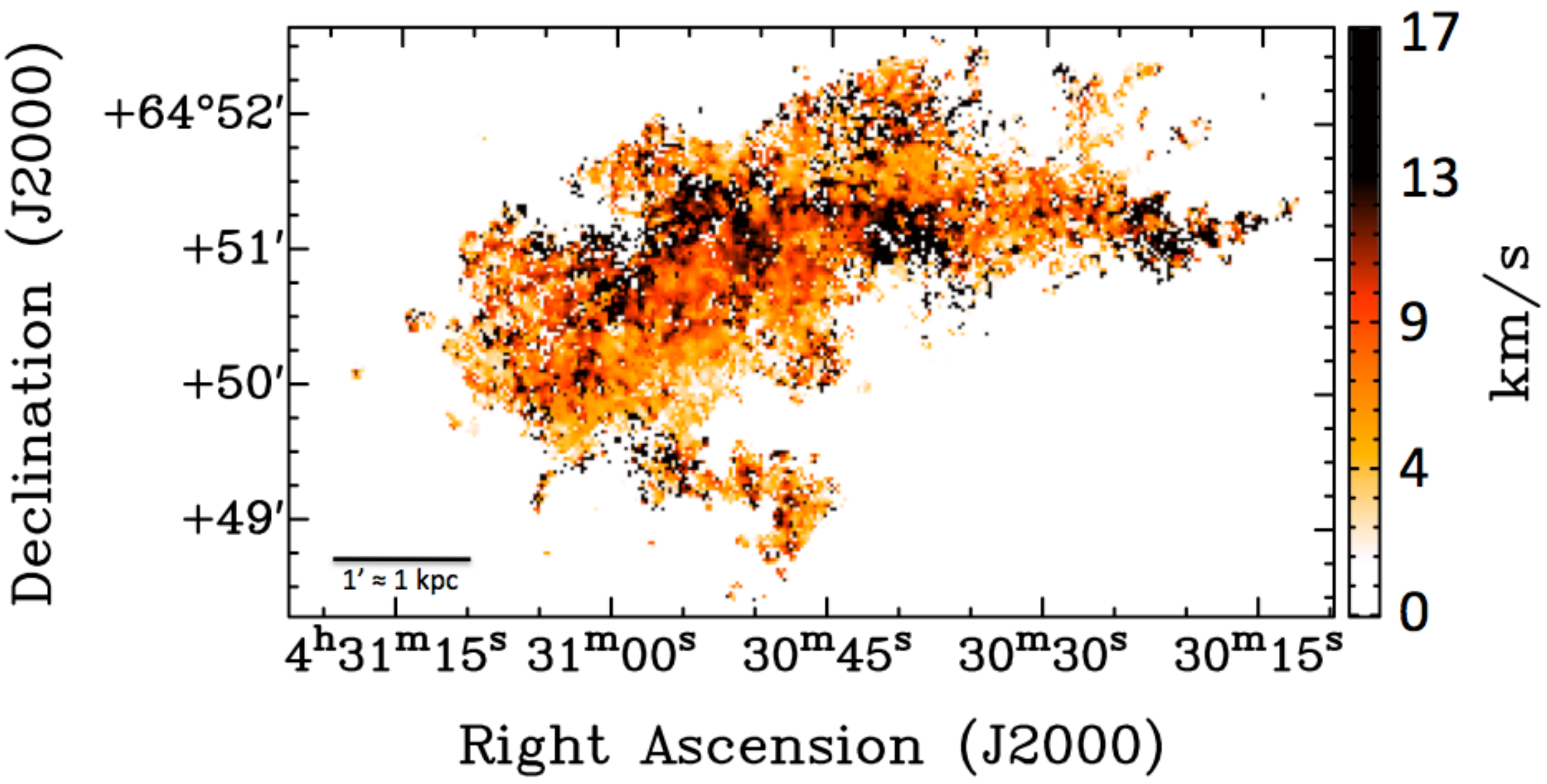}
\caption{Velocity dispersion map of the bulk velocity component only for all extracted velocity peaks above 2$\sigma$.}
\label{fig:59}
\end{figure}

When the stellar velocities are laid on top of an optical image, we can physically see the areas in the galaxy that correspond to observed stellar kinematic features.  
Figure \ref{fig:29} shows the {\it HST} image in greyscale. Red and blue ovals are superimposed on the picture, representing the positions of the extracted spectra along the major axis.  The length of each oval represents the angular area of the slit that we summed, while the width of the ovals represents the 2$\farcs5$ slit width.  
Red and blue ovals represent red- and blue-shifted $V_{\rm helio}$ radial velocities, respectively,  relative to the systemic velocity of NGC 1569.  Each of the ovals are representative of independent data points from spectra taken on different nights.  Therefore, there are some red- and blue-shifted velocities shown at the same positions. The green `X' marks the center of the slit and the center of the galaxy.  

In the region west of SSC A, the overall rotation of the stars is actually red-shifted, which is seen from the dominance of red ovals in Figure \ref{fig:29}.  
The locations of the blue ovals that reside to the west of SSC A correspond to a distance of -27$\arcsec$ from the galaxy center.  If this location is compared with the velocity dispersion plot of the stars for the major axis PA (top left panel) in Figure \ref{fig:20}, the velocity dispersion is larger by $\sim$10 km s$^{-1}$ from the average stellar velocity dispersion.  The increase in stellar velocity dispersion in this region is consistent with secondary star formation that formed out of the surrounding material that was shocked by the expanding shell discussed in Section \ref{sec:4.1.2}.  Therefore, these peculiar kinematics observed in the stars are likely from light 
 composed mostly of young, luminous stars in NGC 1569.  

In the optical {\it HST} image of NGC 1569, shown in the right image of Figure \ref{fig:cartoon}, we can observe \ha\ streamers, which are material likely trapped in the outflow, extending to the northeast and southwest from the SSC region, as indicated by the arrows.  Also, there is a large \ha\ hole where the SSCs reside, also identified in Figure \ref{fig:cartoon}.  We estimate the projected surface area of this hole to be about 0.2 kpc$^2$ at the distance of NGC 1569.  Figure \ref{fig:25} also shows a hole in the \ion{H}{1} in all four PAs.  It is centered around 0$\farcm4$ in the major axis PA and around 0$\farcm1$ in the remaining three PAs in Figure \ref{fig:25} \citep[see][]{isr90}. It is in this hole that we observe the disturbed stellar kinematics. These features are likely evidence of outflow from mechanical energy input from stellar winds and supernova remnants from the collection of massive stars in the SSCs.  Large amounts of X-ray emission coming from this region \citep{ott05a} are consistent with this interpretation.  

\subsection{Shape of NGC 1569}

We explore the three-dimensional shape of NGC 1569 using our kinematic measure, $V_{\rm max}$/$\sigma_{\rm z}$.  For this relationship, we use the asymmetric drift corrected $V_{\rm max}$ from our \ion{H}{1} analysis.  Because we want the maximum velocity of the baryons that are responding to the total gravitational potential in the galaxy, it is necessary to correct for radial pressures in the gas that can artificially mask the true maximum rotation speed.
We derived an asymmetric drift corrected $V_{\rm max}$ of 50 $\pm$ 10 km s$^{-1}$ from the \ion{H}{1} analysis, discussed in Section \ref{sec:3.3.3}, and we use $\sigma_{\rm z}$ of 21 $\pm$ 4 km s$^{-1}$ from the stellar kinematics, discussed in Section \ref{sec:3.1.3}. We follow the same reasoning as \citet{hun05} for our decision to use $\sigma_{\rm z}$ in our $V$/$\sigma$ relation.  We determine a $V_{\rm max}$/$\sigma_{\rm z}$ = 2.4 $\pm$ 0.7.  This is indicative of a thick distribution of young stars that dominate our spectra.  As a reminder, this result assumes that the stars are in a virailized disk and that the Gaussian profiles are symmetric.  Our stellar template spectra span a range of classes and types from F-type main sequence to K-type supergiants and thus, cover a range of stellar ages.  Our observed galaxy spectra are a mix of all the stars in the galaxy but, are dominated by the brightest stars, which are the younger stellar population.  There may be a thin distribution of older stars that our spectra simply were not sensitive to and perhaps, the disk is thicker in the region of SSC A than in the center of the galaxy.
The increase in velocity dispersion and decrease in rotation speed around SSC A are consistent with velocities of stars that formed from shocked gas surrounding the observed expanding shell in the region of SSC A, as discussed above.  Overall, the disk of NGC 1569 appears to be thick compared to spiral galaxies.

 
	\subsection{The Outer Morphology of NGC 1569} \label{sec:5.2}
	
Here, we compare the \ion{H}{1} morphology and velocity fields with our deep H$\alpha$ image of NGC 1569 and the {\it HST} image.  
The left image of Figure \ref{fig:cartoon} shows the contoured integrated \ion{H}{1} map.  This image displays tenuous emission, detected for the first time, extending to the south of the galaxy disk in a teardrop shape, and to the northeast; black ovals outline this emission.  Also identified in this image are two features to the east labeled as `\ion{H}{1} Companion' and `\ion{H}{1} Bridge' identified by \citet{sti98} and \citet{muh05}. Stil \& Israel conclude that these features are indicative of a low mass \ion{H}{1} companion to NGC 1569.  Our data detect this \ion{H}{1} cloud, which can also be easily seen in Figures \ref{fig:47} and \ref{fig:34}
at around $\alpha$ = 04:32:15, $\delta$ = +64:49:00.  The extended \ion{H}{1} emission, circled in Figure \ref{fig:cartoon}, and the two features identified by \citet{sti98} as evidence of a possible low mass \ion{H}{1} companion, may be related to each other and show evidence of a recent interaction or merger with NGC 1569.  We discuss these claims further in Paper II where we present results from deep GBT  \ion{H}{1} mapping around NGC 1569 and three of its nearest neighbors.

Figure \ref{fig:42} shows our deep \ha\ image with the contoured integrated \ion{H}{1} map overlaid so that we can compare the morphologies.  What is interesting here is that there is an UD \ion{H}{1} cloud to the west that is at the edge of the optical galaxy in the same region where we observe the disturbed stellar kinematics.  This \ion{H}{1} cloud is nearly 50 M$_{\sun}$ pc$^{-2}$ more dense than the surrounding medium and if there is molecular gas, the column density could be much larger.  

In Figure \ref{fig:42}, we also observe two smaller, dense \ion{H}{1} clouds on the opposite side, east of the optical galaxy. 
Possibly, there was one continuous, dense \ion{H}{1} structure that connected these two dense \ion{H}{1} clouds with the UD \ion{H}{1} cloud.  The most concentrated stellar light is in between these dense \ion{H}{1} clouds where the SSCs reside, which suggests that the stars in this region formed out of the dense \ion{H}{1} ridge that used to connect these clouds.
The trigger for the starburst could have come from the impact of the in-falling NCM \ion{H}{1} cloud, shown in Figure \ref{fig:13} and identified by the blue oval in Figure \ref{fig:42}, that compressed the interstellar medium of NGC 1569 and caused the dense \ion{H}{1} ridge to collapse. 
It is more likely that the NCM cloud is in-falling into the galaxy, rather than outflowing, because there is evidence of shocked gas near the impact area of the NCM cloud as seen in the continuum maps by \citet[see their Figure 12]{kep10}.       

 We see in Figure \ref{fig:42} that, in general, the \ion{H}{1} density is more concentrated to the north of the SSCs than it is to the south, as shown by the blue arrows and contour levels. 
There is a slight \ion{H}{1} depression in the contour levels right above SSC A as indicated in Figure \ref{fig:42}, possibly created by the outflow of the gas.  This figure also shows the approximate location and projected size of two giant molecular clouds (GMCs) identified by \citet[see Figures 3 and 4]{tay99}. These GMCs lie between the UD \ion{H}{1} cloud and the SSCs and are at the tip of the NCM \ion{H}{1} cloud.  
The region between the UD \ion{H}{1} cloud and the SSCs that contains the GMCs is the most probable area for the next generation of star formation.  \citet{mcq10} suggest that the starburst episode in NGC 1569 is ongoing and started roughly 500 Myr ago.


Also in Figure \ref{fig:42} are streamer-like, arced H$\alpha$ features, which trace HII regions, toward the south and north.  The high density \ion{H}{1} in the north and the visible H$\alpha$ features in the south could be the results of the projection effects from the inclination of the galaxy.  
The X-ray emission is less absorbed to the south, according to \citet{ott05a}, indicating that the north side of the galaxy is pointed toward us, thus closer than the south side, assuming that the emission is escaping in the z-direction.
H$\alpha$ arcs and streamers extend nearly all the way through the \ion{H}{1} disk in the southern direction of the disk.  Similarly, the \ion{H}{1} gas appears more dense in the northern half of the disk because there is more gas along the line-of-sight than there is in the southern half of NGC 1569.

Figure \ref{fig:42} also shows a strong H$\alpha$ chimney to the north \citep{hun93, muh05, wes08}.  The \ion{H}{1} depression noted on Figure \ref{fig:42}, shows the escape route for the H$\alpha$ emission \citep{muh05, wes08}.  Magnetic field lines found by \citet{kep10} may be following the outflow of the gas in this region and are channeling cosmic ray electrons out of the disk.  Although the \ion{H}{1} is more concentrated to the north, due to the projection effect of the inclined disk, the intense stellar winds from the starburst are able to penetrate all the way through the \ion{H}{1} disk in the northern half of the galaxy, too.  Tenuous H$\alpha$ emission to the northeast of the optical galaxy stretches all the way through the \ion{H}{1} disk just as it does in the southern part of the galaxy.  

\citet{wes08} discuss the kinematics of the H$\alpha$ emission and suggest that `blowout,' defined here as gas escaping the gravitational potential of the galaxy, from stellar feedback from the massive stars created in the starburst, is not yet occurring due to decreasing ionized gas velocities at the edges of the expanding H$\alpha$ shells seen near SSC A.  If we assume the total mass of NGC 1569 is equal to our dynamical mass of 1 x 10$^9$ $M_\sun$ (discussed in Section \ref{sec:4.3}), then the escape speed is $\sim$70 km s$^{-1}$ at a radius of 2.2 kpc.  According to Figure \ref{fig:10}, the ionized gas we observe with our \ion{O}{3} spectra show high velocity gas in excess of 70 km s$^{-1}$ indicating that perhaps some of the gas is in the process of being expelled from the galaxy in a blowout. \citet{ott05b} also suggest that NGC 1569 is in the process of losing about 5 x 10$^6$ $M_\sun$ of hot gas, corrected for our distance.  However, Ott et al.\ argue that in order for gas to escape completely from a galaxy, it must be moving at supersonic velocities when it hits the cool surrounding gas at one scale length.  We observe tenuous \ion{H}{1} stretching nearly 12 kpc south of the galaxy that is likely associated with tidal filaments (Paper II), so, the hot gas may cool and fall back onto the disk of NGC 1569 when it hits this cold, neutral gas.  Thus, it is not clear if this hot gas will actually escape the galaxy. 

Figure \ref{fig:35} shows the \ion{H}{1} intensity-weighted velocity contours, with the extended \ion{H}{1} trimmed, superimposed over a low contrast H$\alpha$ image.  In the region around the H$\alpha$ image, we observe a butterfly shape in the \ion{H}{1}. The western wing of the butterfly has velocities that are blue-shifted relative to the systemic velocity, as seen by the blue and purple contours, whereas the eastern butterfly wing has red-shifted velocities, as seen by the orange and red contours and are identified in Figure \ref{fig:35}.  
The location of the UD cloud (solid black oval) and NCM cloud from Figure \ref{fig:13} (dashed black oval) are also indicated.  The velocity structure of the NCM cloud appears different in Figure \ref{fig:35} than in Figure \ref{fig:13} because we are observing the intensity-weighted velocity field instead of the deconvolved velocity structure.
The possible `\ion{H}{1} Companion' and `\ion{H}{1} Bridge' discussed by \citet{sti98} are seen as purple contours and do not follow the rotation of the rest of the galaxy.
Again, we surmise that these features are most likely remnants from a past interaction.  

To the south of the H$\alpha$ image in Figure \ref{fig:35}, we see a cavity in the \ion{H}{1}.  This cavity is likely carved out by ultraviolet photons, winds, and explosions from the massive stars in the starburst.  X-ray emission fills the cavity \citep{mar02, ott05a}.  Furthermore, Figure \ref{fig:35} shows that the spectacular H$\alpha$ tail, known as the `western arm' from Hodge (1974), that extends to the southwest of the optical galaxy is, in fact, the lit up edge of the UD \ion{H}{1} cloud.  This western arm was previously believed to have been associated with the strong NCM cloud created from outflow due to the apparent spatial overlap of the two features \citep{wal91, mar98}.  However, \citet{sti02} argue that the H$\alpha$ tail is, in fact, from infall rather than outflow and that these structures are connected to the \ion{H}{1} bridge.  Similarly, \citet{muh05} argue that \ion{H}{1} gas is being accreted onto NGC 1569 and \citet{ott05a} also see the H$\alpha$ tail emitting soft X-rays.  Infalling \ion{H}{1} gas is a likely scenario, especially since we can see tenuous \ion{H}{1} emission to the south and northeast of NGC 1569, which has the same radial velocities as the \ion{H}{1} bridge.  Because the tenuous \ion{H}{1} gas to the south is spread over such a large region of space, $\sim$92 kpc$^2$, and evidence of interaction is observed in our GBT data, which we report in an upcoming paper, the origin of this tenuous gas, \ion{H}{1} companion, \ion{H}{1} bridge and H$\alpha$ tail coming from outflow from the starburst is unlikely.  We conclude that these features are the result of a recent interaction, which caused the NCM cloud to infall and compress a dense ridge that stretched across the entire optical disk, creating a global starburst which created the SSCs, and left behind what is now the UD \ion{H}{1} cloud.

We see that the velocities of the UD cloud are fairly constant, but, according to Figure \ref{fig:59}, the velocity dispersions at this location are high compared to the average.  Thus, we observe that the UD \ion{H}{1} cloud is kinematically warm because of its high velocity dispersion.  
Perhaps this dense \ion{H}{1} region to the west of the optical disk will be the location of a future star formation episode, provided the gas kinematically cools over time.

\section{Mass Modeling}  \label{sec:4}


	\subsection{Stellar Mass}  \label{sec:4.2}

We use the SED fitting procedure from \citet{zha12} to model the stellar mass and our results are presented in Figure \ref{fig:49}.  Briefly speaking, the Hubble time is logarithmically divided into six independent age bins, and the 
star formation rate (SFR) in each age bin is assumed to be constant. A library consisting of $\sim$ 4 x 10$^6$ different star formation histories (SFHs) is created by 
changing the relative star formation intensity among the six age bins. In addition, the metallicities and internal dust 
extinction are also free parameters when creating the library. The usually adopted two-component (e.g. exponential plus bursts) spectral energy distribution (SED) modeling method can introduce a bias in the
estimation of parameters \citep{zha12}. We use Johnson's U-, B-, and V-band, \ha, 2MASS J-, H-, and Ks-band, and {\it Spitzer} 3.6~$\mu$m photometry data in the SED fit.  In the inner regions (radii $<$ 1.3 kpc) of NGC 1569, we do not use the 3.6~$\mu$m data because the starburst emission in these inner regions is expected to have significant contributions from hot dust \citep{zha10}.  In the outer regions (radii $>$ 1.3 kpc), we use all data except the 2MASS data because they are too shallow to properly model the outer disk.  
Instead of looking for the best-fitting parameter, we use the Bayesian technique to find the 
most probable estimates of parameters by constructing a probability density function and the 
corresponding cumulative distribution function for each parameter, and the confidence interval 
is defined as the central 68\% of the cumulative distribution function. Any possible degeneracies 
between different parameters are reflected in the confidence intervals. 


The right panel of Figure \ref{fig:49} shows the cumulative summed stellar mass as a function of radius.  The profile approaches the horizontal asymptote at a mass of 2.8 x 10$^8$ M$_{\sun}$, giving the total stellar mass of NGC 1569.  
The left panel of Figure \ref{fig:49} shows the surface density mass profile, corrected for inclination.  We see that the stellar mass density profile follows a broken or double exponential.
For these data, we make no assumptions about disk scale height or constant mass-to-light ratio. When we compare a de Vaucouleur's R$^{1/4}$ profile to the surface mass density in Figure \ref{fig:49}, it is clear that it does not fit the data as well as a double exponential profile. 

To see how the R$^{1/4}$ profile compares to the more traditional surface {\it brightness} profile, we plot our 3.6~$\mu$m photometry surface brightness in Figure \ref{fig:57}.  We observe a marginally better fit by the double exponential profile over the R$^{1/4}$ profile.  We see that the R$^{1/4}$ profile overestimates the surface brightness in the center and underestimates it at $\sim$0.5 kpc. 
We conclude that the surface brightness, as well as mass profile of NGC 1569 is best fit as a double or broken exponential disk. 
 
In Figure \ref{fig:57}, there is a break in the exponential profile around 1.3 kpc where it becomes flatter.  This surface brightness profile is a Type III ``antitruncated'' profile, as described by \citet{erw05}. A Type III profile is ascribed to disk flaring in the outer disk in M94 by \citet{her09}.  They attribute this characteristic to a recent interaction or merger.  If this type of
surface brightness profile is generally indicative of a flared disk resulting from an interaction or merger, then this would be additional evidence for an outside influence on NGC 1569.

To check the validity of our derived stellar surface mass density profile, we turn to our observed stellar kinematics.  From \citet[and references therein]{swa99}, we use the relationship between vertical stellar velocity dispersion, $\sigma_z$, and stellar surface mass density, $\Sigma(R)$:
\begin{equation}\label{eqn:5}
\sigma_z = \sqrt{\pi G z_o \Sigma(R)}
\end{equation}
Here, $z_o$ is the vertical scale-height, which we assume is constant in this relationship and is defined as twice the exponential scale height, $z_o$ = 2$h_z$. 
We also assume a vertical $sech^2$ scale-height distribution for the stellar component with the empirical relationship, $h/z_o$ = 5 \citep{van81}, where $h$ and $z_o$ are the disk scale-length and scale-height, respectively.  From an exponential disk fit to the inner part of the 3.6~$\mu$m surface brightness profile, we derive a scale-length, $h$, of 1.54 kpc, giving a scale-height, $z_o$, of 0.31 kpc for NGC 1569.  

Figure \ref{fig:52} shows our observed stellar velocity dispersions, plotted so that all radii are positive, along with the model velocity dispersion profile derived from Equation \ref{eqn:5}.  In Figure \ref{fig:52}, there is general agreement between the observed and predicted stellar velocity dispersions in the center, however, the observed stellar velocity dispersions appear to increase as a function of radius, which is opposite of what is expected.  This anomaly can be explained as an isolated area where the stars have a higher dispersion because they are the stars formed in the expanding shell described in Section \ref{sec:4.1.2}.  These high dispersion stellar velocities coincide with the the peculiar dip, identified by the black arrow, at the edge of the expanding shell in Figure \ref{fig:10} (R = $-$27$\arcsec$ $\Rightarrow$ 0.5 kpc).
We do not know how the stars act beyond $\sim$0.5 kpc, as our spectra do not probe past this radius, so, we cannot definitively determine what happens beyond this increase.

We used the stellar surface density profile, determined from the SED fitting, to model the expected rotation velocities of the stars ({\sc rotmod}), assuming that the total mass of the galaxy comes from the stars.  The results for this model, corrected for inclination, are shown by the asterisk symbols in Figure \ref{fig:50}.  This graph shows that in the inner 1 kpc, half of the observable stellar disk of NGC 1569, the gravitational potential is dominated by the observed stellar mass. 

	\subsection{Gas Mass}  \label{sec:4.1}
	
The total \ion{H}{1} mass of NGC 1569, summed over the integrated \ion{H}{1} map, Figure \ref{fig:47}, is 2.3 x 10$^8$ M$_{\odot}$. 
To calculate the expected velocity of the gas as a function of radius, we use the \ion{H}{1} mass surface density profile.  We assume an axisymmetric, infinitely thin disk comprised entirely of gas and an inclination of 69$\arcdeg$ $\pm\ 7\arcdeg$ \citep[see][for more details]{oh11}. The model \ion{H}{1} velocities as a function of radius are shown by the triangle symbols in Figure \ref{fig:50}. The maximum rotation velocity of the model \ion{H}{1} rotation curve is 18 km s$^{-1}$, less than half of the asymmetric drift corrected $V_{\rm max}$ (filled circles).

NGC 1569 also contains molecular hydrogen (H$_2$).  According to Israel (1988), the amount of H$_2$ in NGC 1569 as a whole is about 5 x 10$^7$ M$_{\odot}$ corrected to a 3.36 Mpc distance.  This produces the ratio $M_{\rm H_2}/M_{\rm HI}$ = 0.2 over the galaxy.  The estimate of the total H$_2$ is an approximation, though, because Israel determined the total H$_2$ mass by scaling the results derived by \citet{you84} for the central 50$\arcsec$ of the galaxy.  Also, the CO to H$_2$ conversion factor used by Israel is potentially dependent on metallicity, and NGC 1569 has a low metallicity. However, the scaled mass agrees with \citet{obr09} who find an H$_2$ mass of $\sim$ 2 x 10$^7$ M$_{\odot}$, corrected for a distance of 3.36 Mpc. In either case, these estimates suggest that the total amount of H$_2$ is on the order of 10\% of the total baryonic mass (stars $+$ gas) in NGC 1569.  

	\subsection{Dark Matter Mass}  \label{sec:4.3}

In the previous sections, we have modeled the contributions to the rotation curve of NGC 1569 due to stars and gas.  Here, we explore the dark matter content in NGC 1569 by assuming that the observed, asymmetric drift corrected, rotation represents motion under the total gravitational potential of the system.  Using our $V_{\rm max}$ = 50 km s$^{-1}$, we calculate a dynamical mass, $M_{\rm dyn}$, of 1.3 x 10$^9$ $M_\sun$ within a radius of 2.2 kpc, using the following equation:
\begin{equation}
M_{\rm dyn}(R) = \frac{V_{\rm max}^2\  R}{G}
\end{equation}
where, $V_{\rm max}$ is the maximum rotation speed determined from the asymmetric drift corrected rotation curve, and $R$ is the radius where $V_{\rm max}$ is measured.
We determine an \ion{H}{1} mass and a stellar mass inside a radius of 2.2 kpc of 1.7 x 10$^8$ M$_\sun$ and 2.5 x 10$^8$ M$_\sun$, respectively.  To account for helium, we scale the \ion{H}{1} mass by a factor of 1.36 \citep{deB08},
and achieve a total baryonic mass of 4.8 x 10$^8$ M$_\sun$.  Subtracting this from the dynamical mass, we are left with only 8.2 x 10$^8$ M$_\sun$ of missing, dark matter within a radius of 2.2 kpc.  

There are, however, a number of uncertainties in the dark matter determination. The most important uncertainty is that we assume that we have found the turnover radius and, hence, the maximum rotation, $V_{\rm max}$, of NGC 1569.  However, the signal-to-noise is low in the region where the turnover radius occurs.  If the true maximum rotation speed of the galaxy is actually higher, then our calculation of dynamical mass, along with the deduced dark matter mass in the central region of NGC 1569, will be underestimated.  Another uncertainty comes from the fact that these calculations use a $V_{\rm max}$ derived under the assumption that the inclination of the \ion{H}{1} disk is 69$\arcdeg$ $\pm$ 7$\arcdeg$, as determined from the tilted ring model.  If a $V_{\rm max}$ of 50 km s$^{-1}$ is correct, then NGC 1569 has a low dark matter content inside the visible disk, which is unlike most other dwarf galaxies studied \citep[e.g.][]{oh11, saw11, wal09}.

To see the effect of such a low dark matter content, we examine the dark matter fraction, $\gamma_{\rm dm} (\frac{M_{\rm DM}}{M_{\rm tot}}$), as a function of radius using the following equation given in \citet{oh11}:
\begin{equation}
\gamma_{\rm dm}(R) = \frac{M_{\rm DM}}{M_{\rm tot}} = \frac{V_{\rm rot}(R)^2 - V_{\rm star}(R)^2 - V_{\rm gas}(R)^2}{V_{\rm rot}(R)^2}
\end{equation}
where $V_{\rm rot}$ is the rotation velocity corrected for asymmetric drift derived from \ion{H}{1}, $V_{\rm star}$ is the stellar rotation velocity assuming all the mass is in the stars determined from the SED fitting, and $V_{\rm gas}$ is the gas rotation velocity determined from the \ion{H}{1} surface density assuming Newtonian dynamics.  Figure \ref{fig:56} shows the fraction of dark matter as a function of radius.  The average values in the inner 1 kpc are zero, within the errors, due to the dominance of the stars.  However, in the outer region, beyond 1.25 kpc, the fraction of dark matter has an average of $\sim$0.5, which is below the average of 0.7 found for a sample of seven dwarf galaxies by Oh et al.  
We see here that the stars account for all of the dynamical mass within a radius of $\sim$1 kpc - no dark matter is needed within this radius.
Thus, NGC 1569 appears to be dark matter deficient in the inner disk compared to other dwarfs.

In $\Lambda$CDM, the dark matter halos hosting galaxies have an approximately universal shape that follows a concentration-mass relation \citep[e.g.][]{bul01, pra12}.
Hence, CDM halos are a one-parameter family, typically parameterized by mass within the virialized extent ($M_{\rm vir}$) or maximum circular velocity ($V_{\rm max}$).  Adding the assumption that there is a monotonic mapping of dark matter halo mass to galaxy luminosity (e.g., the brightest galaxies are hosted by the most massive dark matter halos), a galaxy's dark matter halo can be uniquely determined from its luminosity using the technique of abundance matching 
\citep{kra04,val04,con09}.  
Here, abundance matching is performed, as described in detail in 
\citet{tol11a}, 
using the halo $V_{\rm max}$ function from the Millenium II simulation \citep{boy09} and the $r$ band galaxy luminosity functions of \citet{bla05}.  The $r$ band absolute magnitude for NGC 1569 ($M_r=-17.6$) is determined from the SED fit described in Section \ref{sec:4.2}, and this is used to infer that the halo of NGC 1569 has $V_{\rm max}=95$ km s$^{-1}$.  This implies $M_{\rm vir} \approx 1.6 \times 10^{11} M_{\odot}$, and the virial extent of the halo is $R_{\rm vir} \approx 140$ kpc.

\begin{figure}[H]
\includegraphics[angle=90,scale=.5]{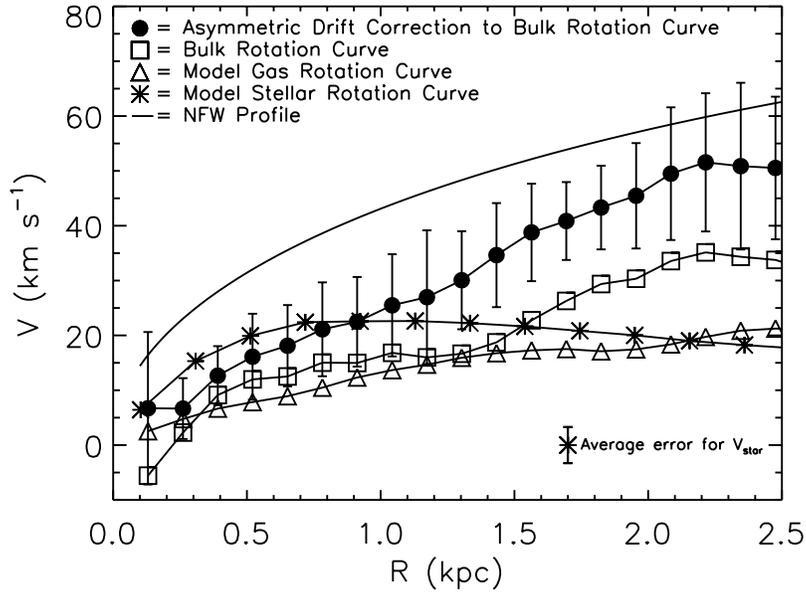}
\caption{Mass modeling of both the stars (asterisks) and the gas (triangles) along with the observed bulk rotation curve (boxes) and the asymmetric drift corrected rotation curve (filled circles).  The NFW profile (solid line) is also shown.  The average uncertainty for the model stellar rotation are shown in the lower right corner for comparison to the asymmetric drift corrected rotation curve.  Within the uncertainties, the stellar mass dominates the entire rotation curve out to 1 kpc, half of the observable stellar and gas disks.}
\label{fig:50}
\end{figure}

\begin{figure}[H]
\includegraphics[scale=.5]{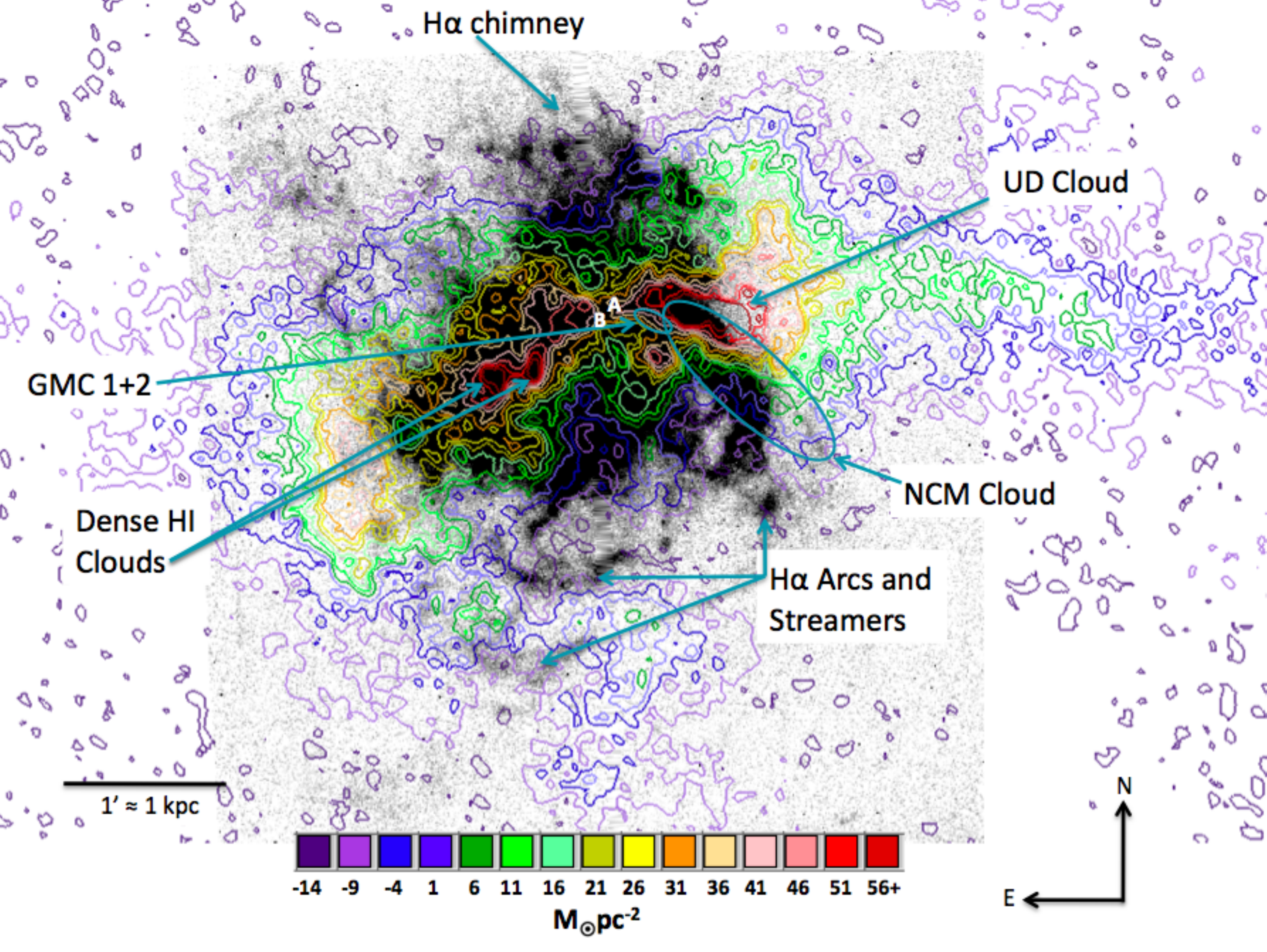}
\caption{We show our deep H$\alpha$ image with the integrated \ion{H}{1} contours over plotted.  More H$\alpha$ streamers and arcs can be seen to the south due to the lower \ion{H}{1} density.  To the north, we observe a strong  H$\alpha$ chimney that extends all the way through the \ion{H}{1} disk. The ultra-dense (UD) \ion{H}{1} cloud is identified as well as the two dense clouds on the opposite side of the optical galaxy.  We believe these features were at one time connected and extended through the current optical galaxy.  The non-circular motion (NCM) cloud, outlined by the blue oval, may have caused the dense \ion{H}{1} gas to collapse into the starburst that ended nearly 20 million years ago.}
\label{fig:42}
\end{figure}

\begin{figure}[H]
\centering
\includegraphics[scale=.4]{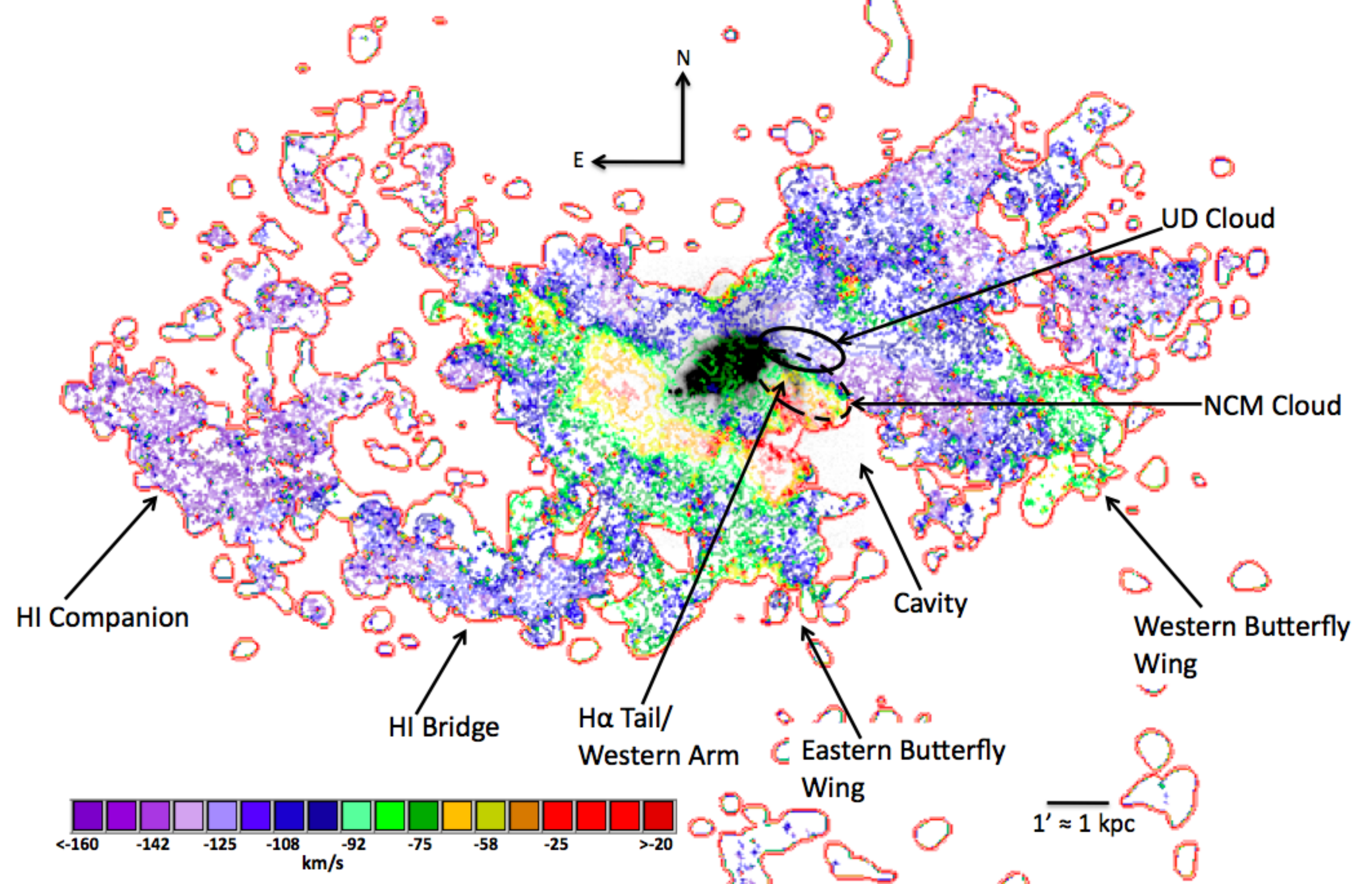}
\caption{Intensity-weighted velocity contour map over-plotted on our H$\alpha$ image.  The features identified by Stil \& Israel (1998) are marked, as well as the position of the NCM cloud shown in Figure \ref{fig:13}.  We show two new features here that make up what we call butterfly wings.  The cavity between the two wings matches the position of X-ray emission \citep{mar02}. We also show the position of the UD \ion{H}{1} cloud.  The H$\alpha$ tail is almost completely within the NCM \ion{H}{1} cloud but has the same velocities as the UD cloud. Therefore, we conclude that this tail is the lit up edge of the UD \ion{H}{1} cloud.}
\label{fig:35}
\end{figure}

\begin{figure}[H]
\includegraphics[angle=90,scale=.5]{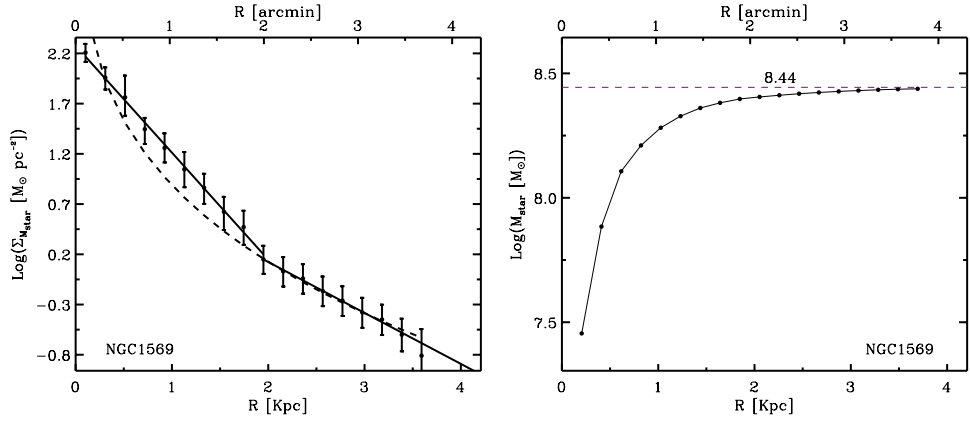}
\caption{\underline{Left}: Logarithmic stellar mass surface density profile from fitting the SED in elliptical annuli.  This has been corrected for an inclination of 69$\arcdeg$ $\pm\ 7\arcdeg$. The solid lines are a fit to the profile and represent a double exponential disk.  The dashed line represents an R$^{1/4}$ fit to the surface density profile where we have assumed a constant mass-to-light ratio and constant disk scale height.  An exponential disk is a better fit than an R$^{1/4}$.  \underline{Right}:  The logarithmic integrated stellar mass for NGC 1569 from the elliptical annuli SED fitting.}
\label{fig:49}
\end{figure}

\begin{figure}[H]
\includegraphics[angle=90,scale=.5]{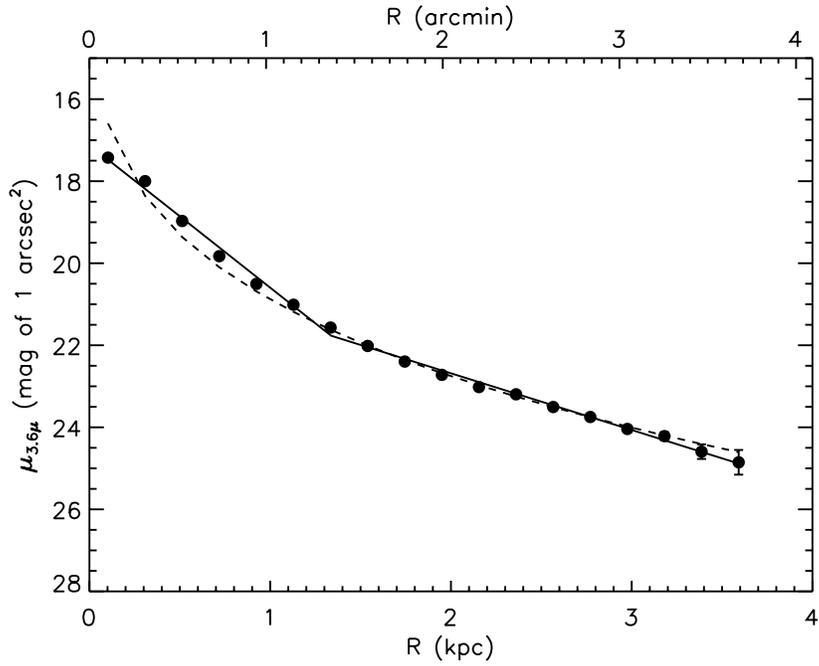}
\caption{Inclination corrected 3.6~$\mu$m surface brightness profile.  The solid lines are an exponential fit to the profile, and the dashed line represents an R$^{1/4}$ fit.  The broken exponential profile fits the data better than the R$^{1/4}$ profile.  We show this figure because it is the traditional way that surface brightness profiles are studied and the transformation to mass surface density, shown in Figure \ref{fig:49}, is not linear.}
\label{fig:57}
\end{figure}

\begin{figure}[H]
\includegraphics[angle=90,scale=.5]{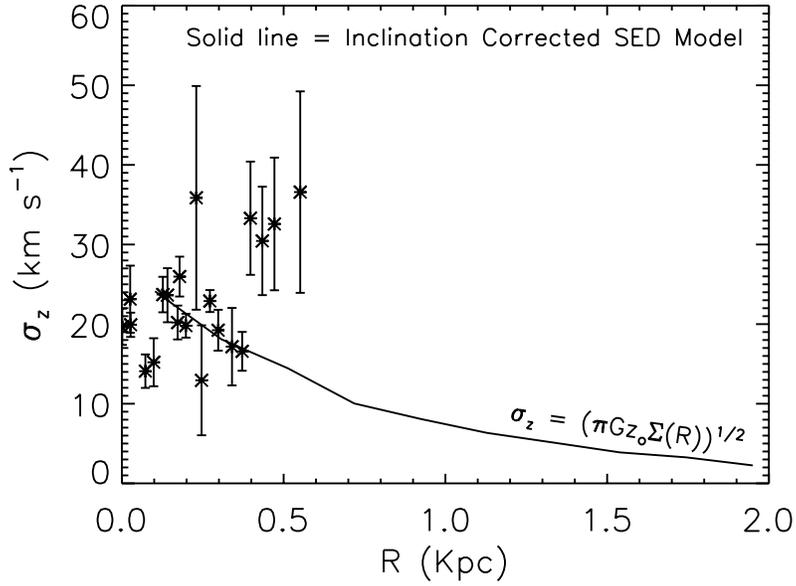}
\caption{Our observed stellar velocity dispersions measured from our spectra (star symbols).  The solid line represents the inclination-corrected model velocity dispersion expected from the stellar mass surface density, determined from SED fitting. $z_o$ is the stellar scale height.  The bulk of the points (R $<$ 0.35 kpc) lie on or near the expected values. But, the points between 0.35 $<$ R $<$ 0.6 kpc are too high, indicating a region of unusually high stellar velocity dispersion.  We interpret this as a disturbed region where the stars formed in an expanding shell.}
\label{fig:52}
\end{figure}

\begin{figure}[H]
\includegraphics[angle=90,scale=.5]{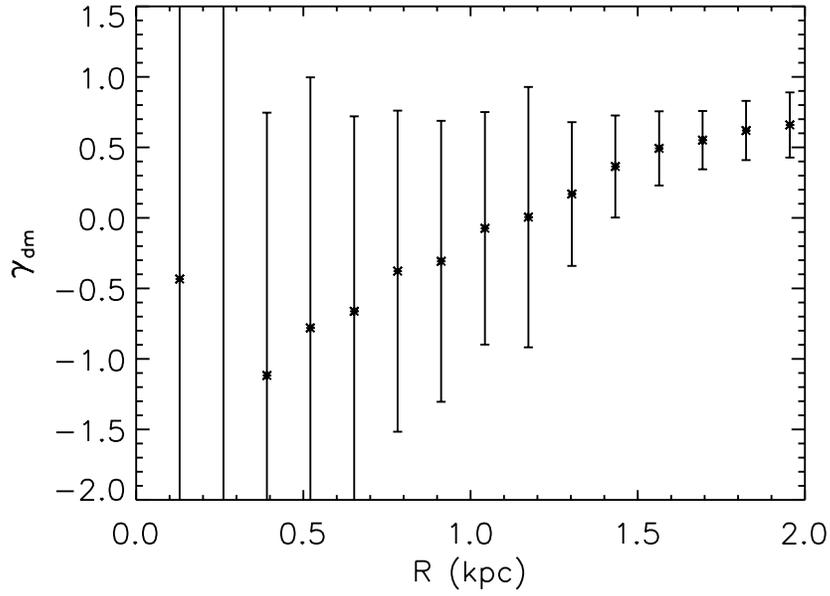}
\caption{Dark matter fraction, $\gamma_{dm}$, as a function of radius.  The negative dark matter fractions in the inner 1 kpc are due to the dominance of the stellar mass to the rotation curve.  Within the errors, the stars account for all of the mass within a radius of 1 kpc. Other dwarfs, by contrast, have $\gamma_{dm}$ $\sim$ 0.7 on average over their entire disks.}
\label{fig:56}
\end{figure}

\begin{figure}[H]
\centering
\includegraphics[scale=.6,angle=90]{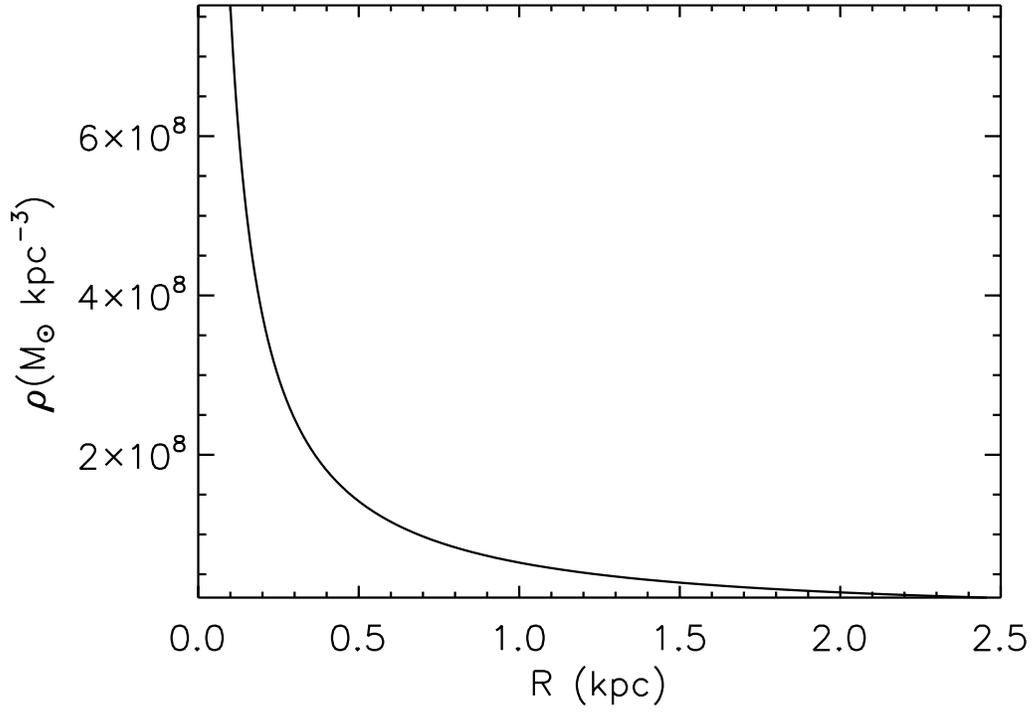}
\caption[NFW density profile]{NFW density profile from the abundance matching method, assuming an $M_r=-17.6$.}
\label{nfw_den}
\end{figure}

\begin{figure}[H]
\centering
\includegraphics[scale=.6]{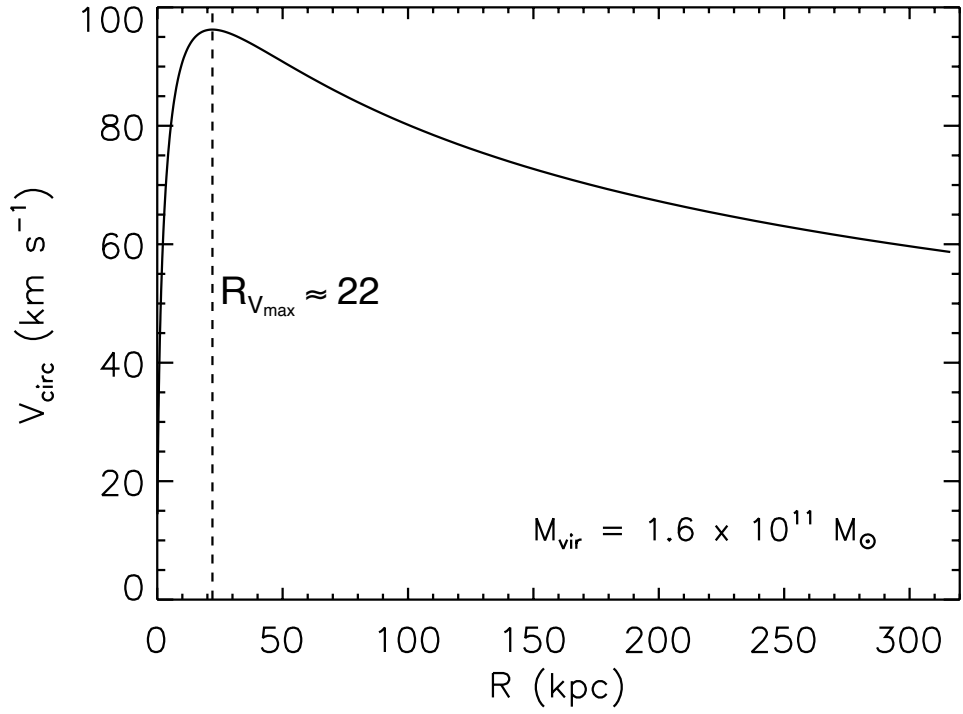}
\caption[NFW velocity profile]{NFW velocity profile for the abundance matching method, assuming an $M_r=-17.6$.}
\label{fig:101}
\end{figure}

With these values for the total mass of NGC 1569's dark matter halo, the concentration-mass relation implies a unique shape for the halo profile.  As outlined in \citet{tol11b}, the halo shape is parameterized using a \citet{nav97} (NFW) halo profile and the shape parameters are assigned so that the halo falls on the median of the concentration-mass relation.  Figure \ref{nfw_den} shows the NFW density profile as a function of radius for NGC 1569.  The resulting circular velocity profile is given as the solid line in Figure \ref{fig:101}.  Figure \ref{fig:50} shows the inner region of the NFW profile, solid line, as compared to the modeled gas and stellar masses.  

As is clear from Figure \ref{fig:50}, the stellar mass is close to the expected dark matter mass in the inner 1 kpc, and the uncertain effect of the baryons on the dark matter halo should be significant here.  Thus, it is ambiguous whether or not the dominance of the baryons in the inner 1 kpc implies a halo profile inconsistent with $\Lambda$CDM, or simply that the baryonic mass of the galaxy is altering the shape of the dark matter halo in the region where the dark matter does not dominate the gravitational forces.

By $R\sim 2$ kpc, however, the $\Lambda$CDM prediction lies within the error bars of the measured rotation curve.  If the model is correct, the hint of a turn over in the rotation curve that we see is likely to be an artifact of low signal-to-noise, as the rotation curve implied by this halo profile does not peak until $R \approx 20$ kpc, far beyond the baryonic extent of the galaxy.  Additionally, it implies that in a global sense, NGC 1569 is not dark matter deficient because at a radius of $\sim$2 kpc, the rotation curve is consistent with $\Lambda$CDM predictions.

\section{Summary}

\subsection{Stellar Kinematics}
\noindent 1)  A systemic velocity, $V_{\rm sys}$, of -82 $\pm$ 7 km s$^{-1}$ was determined from averaging the heliocentric radial velocities of the stars along the minor axis.\\
2) NGC 1569 has a stellar kinematic major axis equal to the morphological major axis at a position angle, measured east from north, of 121$\arcdeg$.\\
3) The stars and gas kinematically follow each other.\\
4) NGC 1569 has an average stellar velocity dispersion of $<\sigma_{\rm z}>$ = 21 $\pm$ 4 km s$^{-1}$ and a kinematic measure, $V_{\rm max}$/$\sigma_{\rm z}$ = 2.4 $\pm$ 0.7, which is suggestive of a thick disk.\\
5) The stellar velocity dispersion increases from 21 $\pm$ 4 km s$^{-1}$ to 33 $\pm$ 9 km s$^{-1}$ in the region of the SSCs at a radius of $\sim$ $+$0.5 kpc from the galaxy center, which is consistent with velocities of stars that formed in shocked gas in the expanding shell around SSC A.  The region in the vicinity of the SSCs is a disturbed part of the disk.\\
6) NGC 1569 follows the observed trend of $\sigma_{\rm z}$ -- $M_{\rm B}$ in large spiral galaxies (Swaters 1999).\\
7) The 3.6~$\mu$m light profile of NGC 1569 follows a double exponential disk better than a de Vaucouleurs's R$^{1/4}$ profile.  The light profile is shallower in the outer disk than in the inner disk.

\subsection{Ionized Gas Kinematics}
\noindent 1) Multiple velocity components are observed in the spectra as a result of mechanical energy input from the starburst.\\
2) There is one expanding shell observed from the high velocity ionized gas seen along the major axis and indications for outflow.  \\
3) The stellar velocities match low velocity ionized and \ion{H}{1} gas and are likely produced by the stars that caused the expanding shell (high velocity gas) observed on the western side of the disk.

\subsection{\ion{H}{1} Kinematics}
\noindent 1) Extended \ion{H}{1} emission to the south and north of NGC 1569 was imaged for the first time.\\
2) NGC 1569 shows a dominance of non-circular motions across the \ion{H}{1} disk as seen in the intensity-weighted velocity field.\\
3) A non-circular motion (NCM) \ion{H}{1} cloud was observed and appears to be impacting NGC 1569 at the eastern edge of a dense \ion{H}{1} cloud that lies west of the SSCs, as determined from a double Gaussian decomposition method that was successfully applied to our \ion{H}{1} data.\\
4) We suggest that the impact from the NCM \ion{H}{1} cloud caused the dense ridge in the center of the disk to become unstable, thus creating the starburst and enabling the formation of the SSCs.

\subsection{Kinematic and Morphological Comparisons}
\noindent 1) The disturbed stellar kinematics correspond to the region west of SSC A. \\
2) A ``butterfly'' shape in the \ion{H}{1} gas morphology is observed, likely caused from outflow from the starburst.  Gas is escaping from the body of the butterfly, perpendicular to the stellar and gas disks.

\subsection{Mass Modeling}
\noindent 1) A total stellar mass of 2.8 x 10$^8$ M$_\odot$ was determined from a new SED fit and a total gas mass, including helium and metals, of 2.3 x 10$^8$ M$_\odot$ was determined.\\
2) A dynamical mass of 1.1 x 10$^9$ $M_\odot$ was calculated from the $V_{\rm max}$ = 50 km s$^{-1}$ at a radius of 2.2 kpc.  \\
3) The stars dominate the gravitational potential in the inner 1 kpc of the \ion{H}{1} disk.\\
4) The derived rotation curve from the \ion{H}{1} kinematics compared to the modeled \ion{H}{1} rotation from the surface mass density and the modeled stellar rotation, shows that only a small amount, 80\% of the baryonic mass, is needed in dark matter.  This is unlike typical dwarf galaxies.\\
5) 
The rotation of the stars and gas seem to agree with an NFW profile at a radius of $\sim$ 2 kpc. \\
6) The NFW model predicts a virial mass of 1.6 x 10$^{11}$ $M_\odot$ and gives a virial radius of 22 kpc, well beyond the radius where baryonic material is detected.

\section{Acknowledgments}

We are grateful to Daryl Willmarth and the KPNO staff for all their help during all of our observing runs. This project was funded by the National Science Foundation under grant numbers AST-0707563 to DAH and AST-0707426 to BGE.  This research has made use of the NASA/IPAC Extragalactic Database (NED) which is operated by the
Jet Propulsion Laboratory, California Institute of Technology, under contract with the
National Aeronautics and Space Administration (NASA).
This research has made use of NASA's Astrophysics Data System and the SAOImage DS9, developed by Smithsonian Astrophysical Observatory.
Parts of this research were conducted by the Australian Research Council Centre of Excellence for All-sky Astrophysics (CAASTRO), through project number CE110001020.

\clearpage

\onecolumn

\clearpage

\begin{deluxetable}{lcr}
\tabletypesize{\small}
\tablenum{1}
\tablecolumns{3}
\tablewidth{0pt}
\tablecaption{Global Parameters for NGC 1569}
\tablehead{
\colhead{Parameter} & \colhead{Value} &\colhead{Ref}
}
\startdata
Other Names & UGC 03056,  ARP 210,& 1\\
	&VII Zw 016 &\\
D (Mpc) & 3.36 $\pm$ 0.20 & 2\\
M$_V$ & -18.2 & 3\\
$\mu_{25}$  (mag arcsec$^{-2}$) & 22.3 & 5\\
Galaxy diameter to 25 mag arcsec$^{-2}$ in $B$, D$_{25}$ (arcmin) & 3.6 & 3\\
$V$-band disk scale length, R$_D$ (arcmin) & 0.39 $\pm$ 0.02 & 3\\
Center (RA, DEC) (J2000) & (04:30:49.8, +64:50:51) & 3\\
Minor-to-major axis ratio, b/a & 0.55 & 3\\
L$_{H\alpha}$ (ergs s$^{-1}$) & 5.7 x 10$^{40}$ & 4\\
Star Formation Rate (SFR$_D$) (M$_{\odot}$yr$^{-1}$kpc$^{-2}$) & 1.3 &4\\
H$_2$ Mass ($M_\sun$) & 5 x 10$^7$ &6\\

\cutinhead{Stellar parameters determined from this work}
$V_{\rm sys}$ (km s$^{-1}$) & -82 $\pm$ 7 & \\
Major Axis PA (degrees) & 121 & \\
Inclination from de-projection, $i_{opt}$ (degrees) & 60 $\pm$ 10 & \\
Average Velocity Dispersion, $<\sigma_{\rm z}>$ (\kms) &21 $\pm$ 4 &\\
Velocity Dispersion near SSC A, $\sigma_{\rm z}$ (\kms) & 33 $\pm$ 9&\\
Total Stellar Mass ($M_\sun$) & 2.8 x 10$^8$ &\\
Optical scale-length, $h$ (kpc)& 1.54&\\
Optical scale-height, $z_o$ (kpc)&0.31&\\

\cutinhead{\ion{H}{1} parameters determined from this work}
$V_{\rm sys}$ (km s$^{-1}$),  & -85  & \\
Major Axis PA (degrees) & 122 & \\
Inclination from tilted ring model, $i_{HI}$ (degrees) & 69 $\pm$ 7 &\\
Kinematic Center ($X_{\rm pos}$, $Y_{\rm pos}$) (RA, DEC)  & (04:30:46.125, +64:51:10.25) & \\
$V_{\rm rot}$ (\kms) & 33 $\pm$ 10&\\
$V_{\rm max}$ (\kms) & 50 $\pm$ 10&\\
Average Velocity Dispersion, $<\sigma_{\rm HI}>$ (\kms) &10  &\\
Total Gas Mass ($M_\sun$) & 2.3 x 10$^8$ &\\
$M_{\rm dyn}$ ($M_\sun$) & 1.3 x 10$^9$ &\\
$M_{\rm DM}$ ($M_\sun$) & 8.2 x 10$^8$ &\\
$M_{\rm vir}$ ($M_\sun$) & 1.6 x 10$^11$ &\\
$R_{\rm vir}$ (kpc) & 140 &\\
$V_{\rm max}$/$\sigma_{\rm z}$ & 2.4 $\pm$ 0.7 &

\enddata
\label{tab:1}
\tablerefs{(1) NASA Extragalactic Database; 
(2) \citet{gro08};
(3) \citet{hun06};
(4) \citet{hun04};
(5) \citet{deV91};
(6) \citet{isr88};
}
\end{deluxetable}

\begin{deluxetable}{rccccccc}
\tabletypesize{\small}
\tablenum{2}
\tablecolumns{8}
\tablewidth{0pt}
\tablecaption{Optical spectroscopic observations of NGC 1569 taken with the KPNO 4-m + Echelle spectrograph in 2008.}
\tablehead{
\colhead{UT Date} & \colhead{PA\tablenotemark{a}} & \colhead{$N_{\rm gal}$\tablenotemark{b}} & \colhead{$T_{\rm gal}$(s)\tablenotemark{c}} & \colhead{$N_{\rm sky}$\tablenotemark{d}} & \colhead{$T_{\rm sky}$(s)\tablenotemark{e}} & \colhead{$N_{\rm spec}$\tablenotemark{f}} & \colhead{$N_{\rm RV}$\tablenotemark{g}}\\
\colhead{Obs.}
}
\startdata
\cutinhead{Mg Ib spectral region}
02/01 & Major axis (121$\arcdeg$) & 5 & 9000 & 5 & 9000 & 7& 4\\
02/02 & Major axis (121$\arcdeg$)&5&9000&5&7393&9&5 $+$ 1\\
02/03 & Minor axis (31$\arcdeg$) &4&7200&4&5673&4&11 $+$ 1\\
11/19 & $+$ 45$\arcdeg$ from major axis (166$\arcdeg$)&8&13,800&7&12,000&4&5\\
11/20 & $-$ 45$\arcdeg$ from major axis (76$\arcdeg$)&3&5400&2&3600&2&6 $+$ 1\\
11/21 & $-$ 45$\arcdeg$ from major axis (76$\arcdeg$)&8&14,400&8&13,860&6&9\\
11/22 & $-$ 45$\arcdeg$ from major axis (76$\arcdeg$)&5&9000&5&8841&3&10\\
&Major axis (121$\arcdeg$)&3&5400&3&5400&4&10\\
\cutinhead{\ion{O}{3} $\lambda$5007 \AA\ observations}
11/23&Major axis (121$\arcdeg$)&3&4800&3&1800& 95&NA\\
11/23&Minor axis (31$\arcdeg$)&3&1800&1&600& &NA\\
\enddata
\label{tab:2}
\tablenotetext{a}{Position angle observed;}
\tablenotetext{b}{Number of galaxy exposures co-added into a single spectrum;}
\tablenotetext{c}{Total integration time, in seconds, on galaxy;}
\tablenotetext{d}{Number of sky exposures co-added into a single spectrum;}
\tablenotetext{e}{Total integration time, in seconds, on sky;}
\tablenotetext{f}{Number of one-dimensional spectra extracted;}
\tablenotetext{g}{Number of radial velocity standard stars used in cross-correlation function.  The ``$+$1'' is in reference to the solar spectrum, from twilight flats, used as a radial velocity standard star. }

\end{deluxetable}

\clearpage

\begin{deluxetable}{lccccccccc}
\tabletypesize{\small}
\tablecaption{Radial velocity standard stars used for cross-correlation.} 
\tablenum{3}
\tablecolumns{11}
\tablewidth{0pt}
\tablehead{\colhead{HD} & \colhead{Apparent} & \colhead{Metallicity} & \colhead{Spectral} & \colhead{RA (J2000)} & \colhead{DEC (J2000)} & \colhead{$V_{\rm helio}$\tablenotemark{a}} & \colhead{$\sigma_{V}$} & \colhead{$T_{\rm exp}$} & \colhead{UT Date} \\
\colhead{No.} & \colhead{Magnitude, V} & \colhead{Fe\_H} & \colhead{Type \& Class} & \colhead{hh:mm:ss} & \colhead{$\arcdeg$ : $\arcmin$ : $\arcsec$} & \colhead{\kms} & \colhead{\kms}  &\colhead{(sec)} &\colhead{Obs. (2008)}}
\startdata
8779&6.41&-0.40\tablenotemark{2}&K0 IV&01:26:53.5&-00:21:18&-5.0&0.5&36&02/03\\
9138 &4.84&-0.37\tablenotemark{3}&K4 III&01:30:37.9&+06:11:15&+35.4&0.5&7&02/03\\
	&&&&&&&&10&11/22\\
12029&7.44&\nodata&K2 III&01:59:11.2&+29:25:16&+38.6&0.5&90&11/22\\
22484 &4.28&-0.10\tablenotemark{3}&F9 IV-V&03:37:18.5&+00:25:41&+27.9&0.1&5 &02/01\\
	&&&&&&&&5&02/02\\
	&&&&&&&&5&02/03\\
	&&&&&&&&5&11/22\\
23169&8.50&\nodata&G2 V&03:44:23.9&+25:45:06&+13.3&0.2&240&02/03\\
26162 & 5.50 & -0.02\tablenotemark{1}&K1 III & 04:09:39.8 & +19:37:52 & +23.9 & 0.6 & 15&02/01\\
32963&7.60&0.08\tablenotemark{1}&G5 IV& 05:08:27.3&+26:20:18&-63.1&0.4&80&02/02\\
	&&&&&&&&80&11/21\\
65583&6.97&-0.56\tablenotemark{3}&G8 V&08:01:03.6&+29:11:09&+12.5&0.4&60&11/19\\
	&&&&&&&&75&11/22\\
65934&7.70&\nodata&G8 III&08:02:42.1&+26:36:49&+35.0&0.3&90&11/22\\
66141&4.39&-0.30\tablenotemark{3}&K2 IIIb Fe-0.5&08:02:42.5&+02:18:39&+70.9&0.3&10&11/22\\
75935&8.46&\nodata&G8 V&08:54:20.3&+26:52:50&-18.9&0.3&170&11/21\\
90861&6.88&\nodata&K2 III&10:30:22.2&+28:32:15&+36.3&0.4&60&11/21\\
92588&6.26&-0.10\tablenotemark{2}&K1 IV&10:41:50.1&-01:47:11&+42.8&0.1&35&11/19\\
	&&&&&&&&40&11/22\\
102494&7.48&-0.26\tablenotemark{4}&G9 IVw...&11:48:22.8&+27:17:36&-22.9&0.3&90&11/22\\
122693&8.11&\nodata&F8 V&14:03:15.6&+24:31:14&-6.3&0.2&170&02/03\\
126053&6.27&-0.45\tablenotemark{3}&G1 V&14:23:41.4&+01:12:08&-18.5&0.4&35&02/02\\
	&&&&&&&&35&02/03\\
132737&7.64&\nodata&K0 III&15:00:14.4&+27:07:37&-24.1&0.3&80&02/01\\
	&&&&&&&&80&02/02\\
	&&&&&&&&80&02/03\\
136202&5.06&-0.08\tablenotemark{3}&F8 III-IV&15:19:44.9&+01:44:01&+53.5&0.2&24&02/03\\
144579 &6.66&-0.69\tablenotemark{1}&G8 IV&16:05:14.4&+39:08:02&-60.0&0.3&35&02/01\\
	&&&&&&&&35&02/03\\
145001&5.00&-0.26\tablenotemark{5}&G5 III&16:08:27.6&+17:01:29&-9.5&0.2&10&02/02\\
	&&&&&&&&10&02/03\\
154417&6.01&-0.04\tablenotemark{6}&F8.5 IV-V&17:05:42.8&+00:41:26&-17.4&0.3&20&02/03\\
182572&5.16&-0.37\tablenotemark{1}&G7 IV H$\delta$ 1&19:25:22.5&+11:57:47&-100.5&0.4&24&11/20\\
	&&&&&&&&24&11/21\\
187691&5.11&0.07\tablenotemark{7}&F8 V&19:51:26.1&+10:26:15&+0.1&0.3&24&11/19\\
	&&&&&&&&24&11/21\\
194071&7.80&\nodata&G8 III&20:22:58.8&+28:16:26&-9.8&0.1&90&11/20\\
203638&5.41&0.02\tablenotemark{5}&K0 III&21:24:38.4&-20:48:55&+21.9&0.1&15&11/20\\
	&&&&&&&&24&11/21\\
212493&4.79&\nodata&K0 III&22:28:17.3&+04:44:19&+54.3&0.3&7&11/19\\
	&&&&&&&&7&11/21\\
	&&&&&&&&10&11/22\\
213014&7.45&\nodata&G9 III&22:28:36.3&+17:18:25&-39.7&0.0&75&11/19\\
213947&6.88&\nodata&K2&22:35:00.6&+26:38:32&+16.7&0.3&40&11/21\\
222368&4.13&-0.15\tablenotemark{3}&F7 V&23:40:23.3&+05:40:21&+5.3&0.2&5&11/20\\
	&&&&&&&&5&11/22\\
223311&6.07&\nodata&K4 III&23:48:58.7&-06:20:00&-20.4&0.1&20&11/20\\
	&&&&&&&&30&11/21\\
\enddata	
\tablenotetext{a}{Heliocentric radial velocities.\\ All data in table, except for metallicity and the last column, come from the United States Nautical Almanac (2008).  Metallicities are listed, where available, and references are given below.}
\tablenotetext{1}{\citet{sou08}} 
\tablenotetext{2}{\citet{ran99}}
\tablenotetext{3}{\citet{cen07}}
\tablenotetext{4}{\citet{yos97}}
\tablenotetext{5}{\citet{mcw90}}
\tablenotetext{6}{\citet{che00}}
\tablenotetext{7}{\citet{fuh98}}
\label{tab:3}
\end{deluxetable}	

\clearpage



\clearpage



\clearpage




\end{document}